\documentclass[fleqn,usenatbib]{mnras}

\pdfoutput=1 
             
\usepackage{newtxtext,newtxmath}

\usepackage[T1]{fontenc}

\DeclareRobustCommand{\VAN}[3]{#2}
\let\VANthebibliography\thebibliography
\def\thebibliography{\DeclareRobustCommand{\VAN}[3]{##3}\VANthebibliography}

\usepackage{graphicx}	
\usepackage{amsmath}	
\usepackage{lscape}
\usepackage{xcolor}
\usepackage{soul}
\usepackage{multicol}
\usepackage{afterpage}
\usepackage{placeins}

\newcommand\tx{\mathrm}

\newcommand{\thb}[1]{\left\langle{#1}\right\rangle}

\newcommand{\gest}{\gamma_{\rm est}}

\defcitealias{Outmezguine_2022}{O23}

\usepackage{lipsum}
\usepackage{enumitem}

\setenumerate[1]{labelsep*=2pt, widest=00.00, align=left, leftmargin=*}

\setenumerate[2]{label=(\alph*), ref=\theenumi{} (\alph*), align=left}

\title[On the Late-Time Evolution of Velocity-Dependent Self-Interacting Dark Matter Halos]{On the Late-Time Evolution of Velocity-Dependent Self-Interacting Dark Matter Halos}

\author[S. Gad-Nasr et al.]{
Sophia Gad-Nasr,$^{1}$\thanks{Email: \href{mailto:sophia.nasr@uci.edu} {sophia.nasr@uci.edu}}
Kimberly K.~Boddy,$^{2}$
Manoj Kaplinghat,$^{1}$
Nadav Joseph Outmezguine,$^{3,4}$
\newauthor
and Laura Sagunski$^{5}$
\\
$^{1}$Center for Cosmology, Department of Physics and Astronomy, University of California - Irvine, Irvine, CA 92697, USA\\
$^{2}$Texas Center for Cosmology and Astroparticle Physics, Weinberg Institute, Department of Physics, The University of Texas at Austin, Austin, TX 78712, USA\\
$^{3}$Berkeley Center for Theoretical Physics, University of California, Berkeley, CA 94720, USA\\
$^{4}$Theory Group, Lawrence Berkeley National Laboratory, Berkeley, CA 94720, USA\\
$^{5}$Institute for Theoretical Physics, Goethe University, 60438 Frankfurt am Main, Germany
}
\date{Accepted XXX. Received YYY; in original form ZZZ}

\pubyear{2023}

\begin{document}
\label{firstpage}
\pagerange{\pageref{firstpage}--\pageref{lastpage}}
\maketitle

\begin{abstract} 
We study the evolution of isolated self-interacting dark matter (SIDM) halos that undergo gravothermal collapse and are driven deep into the short-mean-free-path regime. 
We assume spherical Navarro-Frenk-White (NFW) halos as initial conditions and allow for elastic dark matter self-interactions. 
We discuss the structure of the halo core deep in the core-collapsed regime and how it depends on the particle physics properties of dark matter, in particular, the velocity dependence of the self-interaction cross section. We find an approximate universality deep in this regime that allows us to connect the evolution in the short- and long-mean-free-path regimes, and approximately map the velocity-dependent self-interaction cross sections to constant ones for the full gravothermal evolution. We provide a semi-analytic prescription based on our numerical results for halo evolution deep in the core-collapsed regime.
Our results are essential for estimating the masses of the black holes that are likely to be left in the core of SIDM halos.

\end{abstract}




\section{Introduction}
The nature of dark matter remains one of the most pressing questions in physics. We have a wealth of data that constrains the interactions of dark matter with the Standard Model (SM). However, less is known about the possible self-interactions of dark matter. Self-interactions 
arise generically in dark sector models of physics beyond the SM and astrophysics provides a way to constrain or measure its strength~\citep{Tulin_2018,Adhikari:https://doi.org/10.48550/arxiv.2207.10638}. 
Early models of SIDM featured contact interactions with constant cross sections~\citep{Spergel_2000}, but it has become evident that the cross section must vary with velocity in order to match observations of cores in dwarf galaxies~\citep{Elbert_2015,Tulin_2018,Turner:2021,Adhikari:https://doi.org/10.48550/arxiv.2207.10638} while avoiding constraints on cluster scales~\citep{Buckley_2010PhRvD..81h3522B,Loeb_2011PhRvL.106q1302L,Peter_2013MNRAS.430..105P,Kaplinghat:2015aga,Elbert_2018ApJ...853..109E,Tulin_2018,Sagunski_2021}. Viable velocity-dependent SIDM models also lead to a large diversity in the rotation curves of galaxies as observed~\citep{Oman_2015} while
maintaining all the successes of $\Lambda$CDM on large scales~\citep{Kamada_2017,Ren:2018jpt}.

A key feature of the evolution of SIDM halos is the possibility of gravothermal core collapse~\citep{Balberg:2002ue}. Core collapse is unlikely to occur in the vast majority of field halos, however the gravothermal evolution will be faster in some satellite galaxies and subhalos, which could result in core collapse~\citep{Nishikawa_2020,Sameie:2019zfo,Zeng:2021ldo,Yang:2022mxl}. It has been argued that core collapse must occur in some of the satellite galaxies if SIDM is to explain the diversity in the halo densities of the Milky Way (MW) satellites~\citep{Kahlhoefer:2019,Kaplinghat:2019svz,Zavala:2019,Correa:2021,jiang:2021,Turner:2021}. It seems clear from recent work comparing the densities of the MW satellites to SIDM models that do not allow for collapse that such models cannot be viable~\citep{Read:2018MNRAS.481..860R,Valli:2018,Correa:2021,Kim:2021zzw,Silverman:2023MNRAS.518.2418S}. 
If the gravothermal collapse proceeds far enough in some halos, we expect the mean free path of dark matter particles to become significantly shorter than the core size. In this regime, we do not have accurate simulations or analytic framework. In this work, we explore the gravothermal core collapse phase in the short mean free path regime allowing us to extend existing analytic approximations to high core temperatures. 

\begin{table*}
\centering
\begin{tabular}{|| c l ||} 
 \hline
Parameter & Definition\\ [0.5ex]  
 \hline\hline
 $r_{\rm core}$ & Radius at which the relation $\rho(r_{\rm core})=\rho_c/2$ is satisfied for each snapshot in time \\
 \hline
 $t_{\rm core}$ & Time at which the core size is a maximum (and central density is a minimum) \\
 \hline
 $t_{c,0}$  &  LMFP scattering timescale (see Eq.~\eqref{eq:t_LMFP}) \\
 \hline\hline
$X_c(t)$  & Variables that describe the properties of the core (densest region in the halo) like the mass ($M_c(t)$),\\ & density ($\rho_c(t)$) and velocity dispersion ($v_c(t)$)  \\
 \hline
 $X_{c,0}\equiv X_c(t_{\rm core})$ & Core properties at the time $t_{\rm core}$ \\
 \hline
 $X_{c,\rm LS}\equiv X_c(t_{\rm LS})$ & Core properties at the time when the scatterings in the core transition from being in the long (LMFP) 
 \\ & to short mean-free-path (SMFP) regime \\
 \hline
 $X_{c,10}\equiv X_c(t_{10})$ & Core properties at the time the core enters a new phase deep in the SMFP (Stage 3 in \S\ref{stages}) \\
 \hline\hline
 $X_N$  & Scale parameters used to make the gravothermal equations dimensionless  \\
 \hline
 $\sigma_{c,0},\,\sigma_{c,\rm LS},\,\sigma_{c,10}$  &  Value of the cross section, $\sigma_0K_p$, with the velocity dependence taken at $v_{c,0},\,v_{c,\rm LS},\,v_{c,10}$ (see Eq.~\eqref{eq:K_p}) \\
 \hline
 $n_{c,0},\,n_{c,\rm LS},\,n_{c,10}$ & Log-slope of the cross section at the velocity scales $v_{c,0},\,v_{c,\rm LS},\,v_{c,10}$ (see Eq.~\eqref{eq:n_func}) \\
 \hline
 $\hat{\sigma}_{c,0},\,\hat{w}_{c,0}$ & $\hat{\sigma}$ and $\hat{w}$ in Eq.~\eqref{eq:sigma_hat} evaluated at the scattering timescale $t_{c,0}$\\
 \hline
\end{tabular}
\caption{Table of parameters and their definitions. The first block of parameters are the core radius $r_{\rm core}$, the time of core formation $t_{\rm core}$, and the scattering timescale $t_{c,0}$. In the second block, the variable $X$ is a placeholder for all the physical quantities used in the paper. The quantities $\rho_c,\,v_c$ are taken at the center of the halo (see Appendix~\ref{sec:Core_resolution_test} for details about how we address this numerically), while $r_c,\,M_c,\,L_c$ are taken at the core radius, which is definition dependent. In the last block, we include definitions involving the scattering parameters $\sigma_c,\,n_c,\,\hat{\sigma},\hat{w}$.}
\label{table:param_definition}
\end{table*}

The evolution of an isolated, thermalized SIDM halo can be described by a set of gravothermal fluid equations.
In the early stages of evolution, the halo is sufficiently dilute such that a dark matter particle completes many orbits before scattering with another dark matter particle.
In this long-mean-free-path (LMFP) regime, the gravothermal evolution exhibits self-similar behavior, as demonstrated in our previous work (\cite{Outmezguine_2022}, hereafter referred to as \citetalias{Outmezguine_2022}) and other works (for example,~\cite{LyndenBell:1980,Balberg:2002ue,Koda:2011yb}).
The gravothermal evolution is unstable, leading to a runaway process in which the core of a halo collapses~\citep{Lynden-Bell:1968eqn,Kochanek_2000ApJ...543..514K,Balberg:2002ue,Colin_2002ApJ...581..777C,Koda:2011yb,Pollack:2014rja,Elbert_2015,Sameie_2018MNRAS.479..359S,Essig:2018pzq,Kahlhoefer:2019,Zavala:2019,Nishikawa_2020,Correa:2021,Turner:2021}.
Due to the self-similarity, all gravothermal halos undergo universal evolution leading to core collapse, and the physical timescale on which the collapse occurs depends on the particle physics of the halo, such as the SIDM cross section \citep{Balberg:2002ue,Outmezguine_2022}, truncation of the NFW profile due to tidal effects~\citep{Nishikawa_2020,Correa:2021}, the presence of baryons~\citep{Robles_2019MNRAS.490.2117R,Sameie_2018MNRAS.479..359S,Feng_2021}, and dissipative interactions \citep{Essig:2018pzq,Xiao:2021JCAP...07..039X,Feng_2022JCAP...05..036F}.

As gravothermal evolution proceeds, the core of the halo is driven deep into the core-collapse phase, entering the short-mean-free path (SMFP) regime in which the high density of particles allows them to undergo numerous collisions along each orbit. The evolution of the core in the SMFP regime becomes distinct from that of the outer halo, which remains in the LMFP regime. The properties of the core change much more rapidly in the SMFP regime than in the LMFP regime; for example, the central density of the core increases many orders of magnitude in extremely short time periods, and the temperature increases much more rapidly, compared to the LMFP evolution. Thus, the universality of the core evolution in the LMFP regime is not expected to hold in the SMFP regime.
No previous studies have investigated the SMFP regime to see if the core maintains some form of universal evolution, even if approximate.
If such a universality exists, the gravothermal evolution of any halo could be generically characterized from its initial stage to its gravothermal catastrophe, when the core is expected to reach a relativistic instability that produces a black hole.

Previous works have estimated black hole masses resulting from the relativistic instability (see for example~\citet{Balberg:2002ue,Koda:2011yb,Nishikawa_2020,Meshveliani_2022}). However, lacking accurate evolution characteristics of halos in the SMFP regime results in inaccurate estimates of the core mass at the relativistic instability. If black holes can be produced from the core collapse of dark matter halos, production always occurs deep in the SMFP regime. It is therefore imperative to understand the SMFP core evolution to determine the core properties in this regime.

In this paper, we explore the SMFP regime and study the evolution of the cores of initially NFW, isolated, virialized SIDM halos by solving the spherically symmetric gravothermal fluid equations.
We show that the universality exhibited in \citetalias{Outmezguine_2022} breaks down when the halo transitions from the LMFP to the SMFP regime.
We discover a new, approximately universal solution to the gravothermal equations during the phase of evolution when the thermal energy of the core becomes constant with respect to the 1-dimensional (1D) velocity dispersion and time.
The transition into this phase cannot be predicted analytically. Fortunately, we find that the transition point can be related to the LMFP-to-SMFP transition, which is analytically described in \citetalias{Outmezguine_2022}. We also 
quantify the scaling of the core mass as a function of the central velocity dispersion during this phase of approximate SMFP universality. 
We find more accurate relations for the core structure in the SMFP. As an application, we find the core mass at the relativistic instability, which would be the minimum mass available for black hole formation, and show that the mass loss is steeper than previously thought, resulting in smaller core masses at the relativistic instability. Finally, we present a recipe to easily compute the core properties of halos evolving deep in the SMFP regime.

This paper is organized as follows:
In \S\ref{gravothermal_evo} we present the gravothermal equations and expressions for the heat conductivity, and we summarize our numerical procedure.
In \S\ref{stages} we discuss the LMFP and SMFP evolution and formulate new designations for each stage of evolution.
In \S\ref{SMFP_universality} we present the analytic description of the new SMFP universality and validate it with our numerical solutions of the gravothermal equations in the SMFP regime. 
In \S\ref{black_hole_estimates} we implement a semi-analytic method to determine the parameters in the constant-thermal-energy phase and outline a step-by-step recipe for obtaining the SMFP core mass and black hole mass. The implications and limitations of this work are summarized in the conclusions in \S\ref{sec:conclusions}. We include a derivation of the SMFP universal solution in~Appendix~\ref{appendix_derivations}, show the dependence of the new SMFP parameters on the concentration of the halo in~Appendix~\ref{concentration_dep}, address issues of numerical resolution in Appendix~\ref{resolution}, and discuss the validity of approximations used in this paper in Appendix~\ref{appendix_approx}.

\section{Gravothermal Evolution}
\label{gravothermal_evo}
The gravothermal evolution for a spherically symmetric, isolated, virialized halo is given by the following relations for mass conservation, hydrostatic equilibrium, Fourier's Law, and the first law of thermodynamics~\citep{LyndenBell:1980,1987degc.book.....S,Balberg:2002ue,Nishikawa_2020}:
\begin{align}\label{eq:gravothermal}
	\frac{\partial M}{\partial r}&=4\pi r^2\rho\;\;, \;\;\frac{\partial(\rho v^2)}{\partial r}=-\frac{GM\rho}{r^2}\;\;,\;\;\frac{L}{4\pi r^2}=-\kappa\frac{\partial T}{\partial r}\;\;,\nonumber\\
	\frac{\partial L}{\partial r}&=-4\pi r^2\rho v^2\left(\frac{\partial}{\partial t}\right)_M \mathrm{log}\left(\frac{v^3}{\rho}\right),
\end{align}
where $\rho$ is the halo mass density, $M$ is the mass enclosed within radius $r$, $L$ is the luminosity, and $v$ is the 1D velocity dispersion. The entropy appears in the first law as $\log(v^3/\rho)$, and the temperature is related to the 1D velocity dispersion via $T=m_{\rm dm}v^2$ from the kinetic theory of gases, where $m_{\rm dm}$ is the dark matter particle mass.

\begin{table*}
\addtolength{\tabcolsep}{-2.3pt}
\centering
\makebox[\textwidth]{\begin{tabular}{|| c c c c c c c c c c c c c c c c||} 
 \hline
Run & Color & $\frac{\sigma_{0}}{m_\textrm{dm}}$ & $w$ & $\rho_{c,\tx{LS}}$ & $v_{c,\tx{LS}}$ & $\rho_{c,10}$ & $v_{c,10}$ & $\frac{\sigma_{c,0}}{m_\textrm{dm}}$ & $\frac{\sigma_{c,\tx{LS}}}{m_\textrm{dm}}$ & $\frac{\sigma_{c,10}}{m_\textrm{dm}}$ & $n_{c,0}$ & $n_{c,\tx{LS}}$ & $n_{c,10}$ & $\hat{\sigma}_{c,0}$ & $\hat{w}_{c,0}$ \\ 

  & & $[\frac{\textrm{cm}^2}{\textrm{g}}]$ & $[\frac{\textrm{km}}{\textrm{s}}]$ &  $[10^{10} \frac{M_\odot}{\textrm{kpc}^3}]$ & $[\frac{\textrm{km}}{\textrm{s}}]$ & $[10^{14} \frac{M_\odot}{\textrm{kpc}^3}]$ & $[\frac{\textrm{km}}{\textrm{s}}]$ & $[\frac{\textrm{cm}^2}{\textrm{g}}]$ & $[\frac{\textrm{cm}^2}{\textrm{g}}]$ & $[\frac{\textrm{cm}^2}{\textrm{g}}]$ &  &  & & &  \\ [0.5ex]  
 
 \hline\hline
1 & \textcolor[HTML]{08306b}{\textbf{------}} & $5.0$ & $10^4$ & 3.3 & 39.5 & 3.9 & 100.2 & 5.0 & 4.89 & 4.88 & $2.8\times10^{-4}$ & $4.6\times10^{-4}$ & $3.0\times10^{-3}$ & 0.03 & 339 \\
 \hline
 2 & \textcolor[HTML]{08519c}{\textbf{------}} & $5.5$ & 535 & 3.3 & 39.5 & 2.8 & 93.9 & 5.2 & 4.98 & 3.76 & 0.09 & 0.15 & 0.58 & 0.03 & 18.2  \\
 \hline
 3 & \textcolor[HTML]{2171b5}{\textbf{------}} & $6$ & 298 & 3.7 & 39.8 & 1.9 & 86.3 & 5.2 & 4.68 & 2.74 & 0.26 & 0.39 & 1.0 & 0.03 & 10.1 \\
 \hline
 4 & \textcolor[HTML]{4292c6}{\textbf{------}} & $7$ & 184.7 & 4.5 & 40.2 & 2.0 & 83.3 & 5.1 & 4.14 & 1.81 & 0.53 & 0.77 & 1.52 & 0.03 & 6.3 \\
 \hline
 5 & \textcolor[HTML]{6baed6}{\textbf{------}} & $11$ & 103.9 & 6.0 & 40.9 & 3.9 & 85.2 & 5.1 & 3.48 & 0.94 & 1.02 & 1.37 & 2.18 & 0.03 & 3.5 \\
 \hline
 6 & \textcolor[HTML]{9ecae1}{\textbf{------}} & $42$ & 41.7 & 14.5 & 42.9 & 21.6 & 93.4& 5.0 & 2.13 & 0.26 & 1.97 & 2.4 & 3.0 & 0.03 & 1.4  \\
 \hline
 7 & \textcolor[HTML]{c6dbef}{\textbf{------}} & $1050$ & 11.8 & 40.0 & 45.5 & 144.2 & 104.5 & 5.1 & 1.15 & 0.07 & 3.01 & 3.3 & 3.5 & 0.03 & 0.4 \\
 \hline
 8 & \textcolor[HTML]{deebf7}{\textbf{------}} & $5\times10^6$ & 1.0 & 92.3 & 47.7 & 666.7 & 115.3 & 5.0 & 0.71 & 0.03 & 3.67 & 3.7 & 3.8 & 0.03 & 0.03 \\
 \hline\hline
 9 & \textcolor[HTML]{7f2704}{\textbf{------}} & 90 & $10^4$ & 0.016 & 31.4 & 0.016 & 77.4 & 90.0 & 88 & 87.9 & $2.8\times10^{-4}$ & $2.9\times10^{-4}$ & $1.8\times10^{-3}$ & 0.5 & 339 \\ 
 \hline 
 10 & \textcolor[HTML]{a63603}{\textbf{------}} & 96 & 535 & 0.016 & 31.4 & 0.011 & 72.6 & 91.6 & 89.4 & 74.5 & 0.09 & 0.1 & 0.4 & 0.5 & 18.2\\ 
 \hline
 11 & \textcolor[HTML]{d94801}{\textbf{------}} & $105$ & 298 & 0.017 & 31.5 & 0.009 & 68.8 & 91.2 & 88.5 & 59.3 & 0.26 & 0.27 & 0.82 & 0.5 & 10.1 \\  
 \hline 
 12 & \textcolor[HTML]{f16913}{\textbf{------}} & $125.4$ & 184.7 & 0.017 & 31.5 & 0.009 & 67.2 & 90.8 & 87.2 & 44.0 & 0.53 & 0.56 & 1.29 & 0.5 & 6.3 \\
 \hline 
 13 & \textcolor[HTML]{fd8d3c}{\textbf{------}} & 195 & 103.9 & 0.017 & 31.6 & 0.01 & 65.6 & 91.1 & 84.7 & 28.5 & 1.02 & 1.09 & 1.9 & 0.5 & 3.5 \\ 
 \hline 
 14 & \textcolor[HTML]{fdae6b}{\textbf{------}} & 770 & 41.7 & 0.021 & 31.9 & 0.02 & 67.3 & 92.3 & 76.2 & 12.1 & 1.97 & 2.1 & 2.8 & 0.5 & 1.4\\ 
 \hline
 15 & \textcolor[HTML]{fdd0a2}{\textbf{------}} & $1.9\times10^4$ & 11.8 & 0.029 & 32.6 & 0.09 & 73.8 & 92.0 & 60.8 & 4.12 & 3.01 & 3.1 & 3.4 & 0.5 & 0.4 \\  
 \hline 
 16 & \textcolor[HTML]{fee6ce}{\textbf{------}} & $9\times10^7$ & 1.0 & 0.041 & 33.2 & 0.44 & 83.3 & 90.8 & 48.8 & 1.58 & 3.67 & 3.7 & 3.8 & 0.5 & 0.03 \\
 \hline
\end{tabular}}
\caption{
Table of parameters for all halo runs. Input parameters for the code are $\rho_s,\, r_s$, $\sigma_0/m_{\rm dm}$, and $w$. 
The parameters $\rho_{c,\rm LS},\,v_{c,\rm LS},\,\rho_{c,10},\,v_{c,10}$ are obtained numerically, as are the parameters at core formation: $\rho_{c,0} \simeq2.4\rho_s = 4.8\times10^7\,M_\odot/{\rm kpc}^{3}$, $v_{c,0} \simeq0.64V_{\rm max} =29.5\,\rm km/{\rm s}$.
All remaining parameters are derived. 
Runs $1-8$ all have $\sigma_{c,0}\approx 5\,\mathrm{cm}^2/\mathrm{g}$, while runs $9-16$ have $\sigma_{c,0}\approx 90\,\mathrm{cm}^2/\mathrm{g}$. All runs have $\rho_s = 2\times10^7$ $M_\odot/\textrm{kpc}^{3}$, $r_s = 3\,{\rm kpc}$, $V_{\rm max}\approx 1.65r_s\sqrt{G\rho_s} =45.9\,\rm km/{\rm s}$. All $\hat{\sigma}_{c,\rm LS}=1$ by definition. See Table~\ref{table:param_definition} for the explanation of our notation.
}
\label{table:run_params}
\end{table*}

\begin{figure*}
\centering
\includegraphics[width=0.49\textwidth]{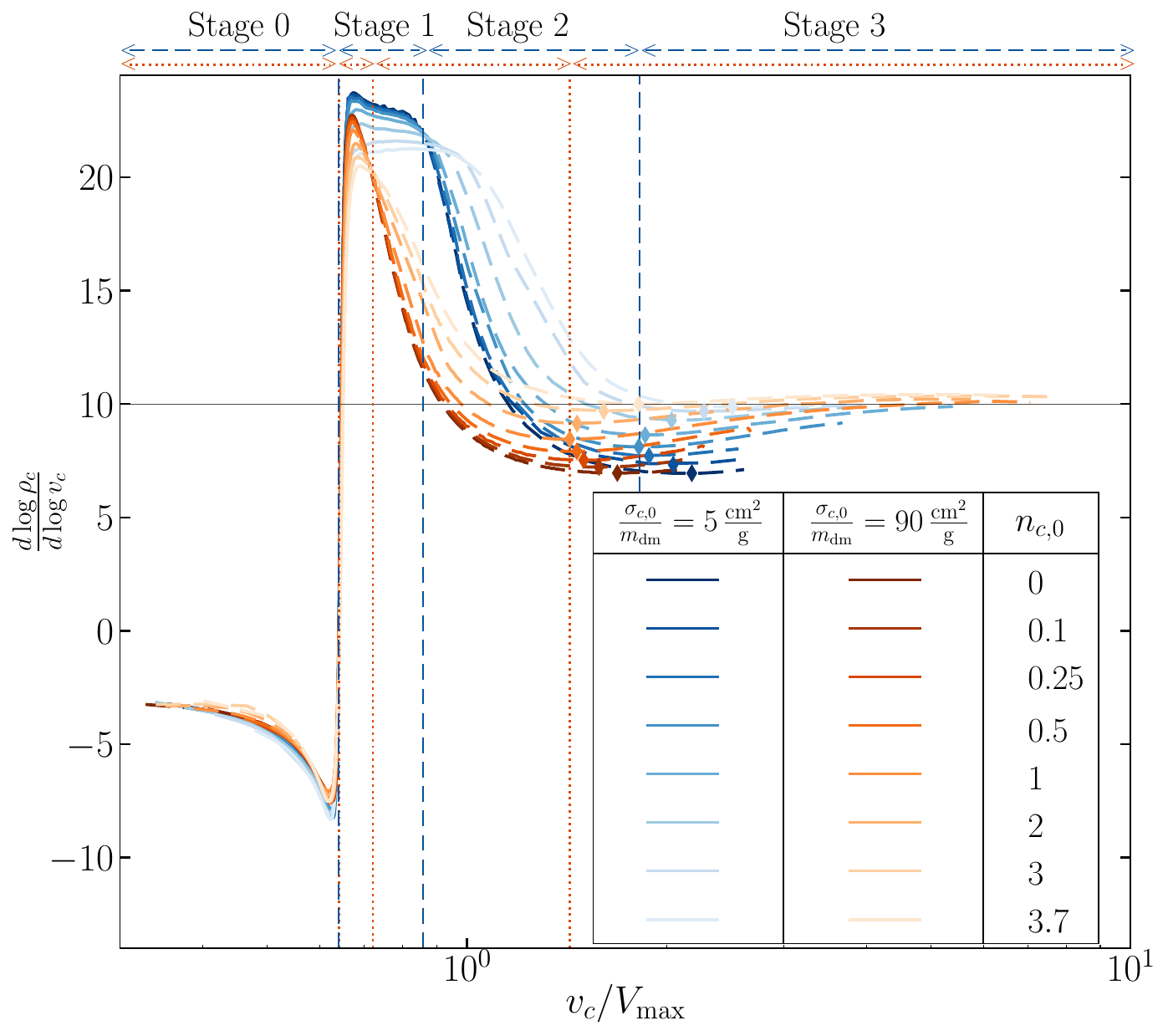}\;\;\;\;\includegraphics[width=0.48\textwidth]{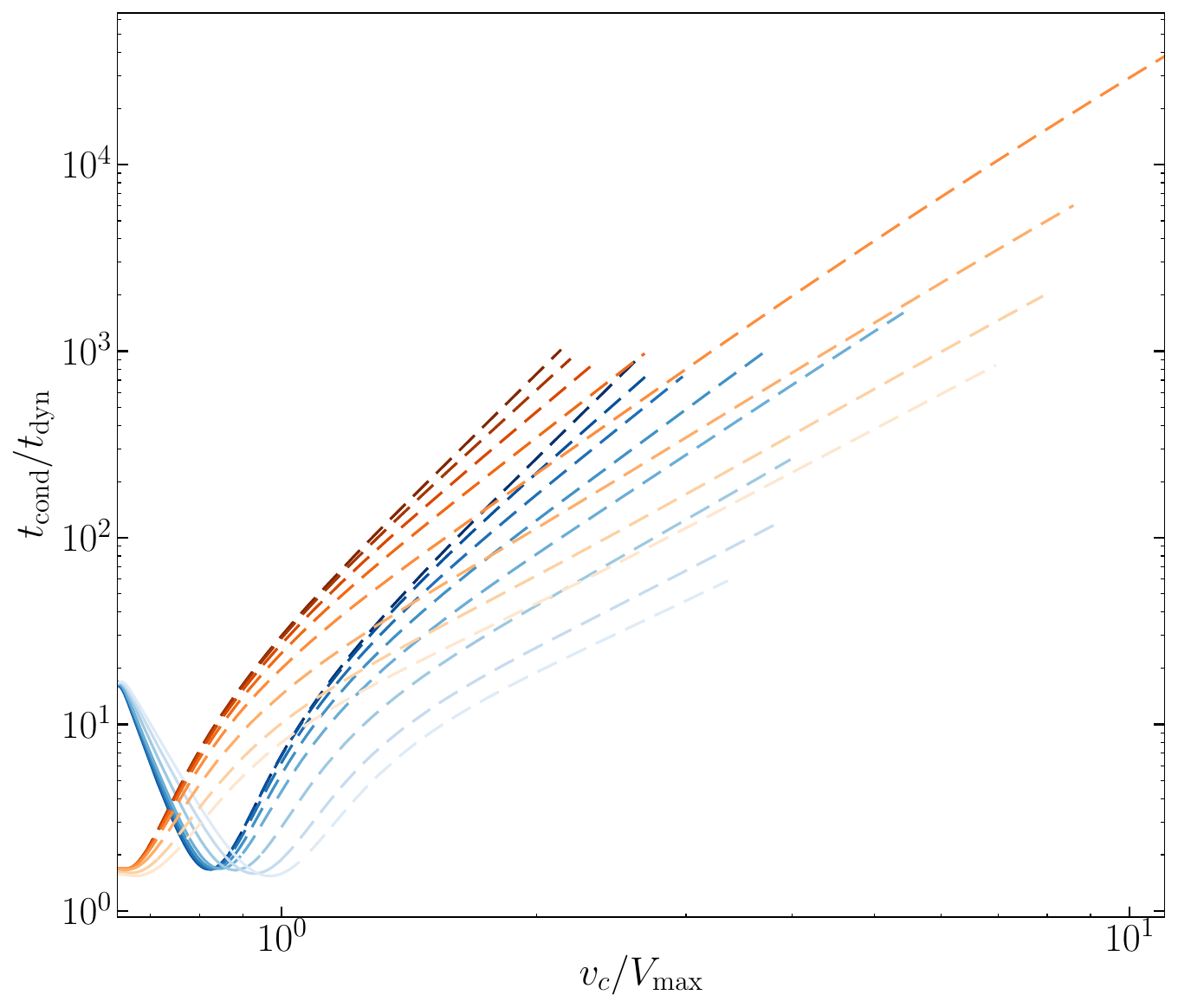}
\caption{ 
{\it(Left)} The log slope of the central density as a function of the central velocity, normalized by $V_{\rm max}$. Dashed lines represent the SMFP evolution of each halo. The black horizontal line shows the $\gamma=10$ slope the halos asymptote to after the $\gamma=10$ transition (indicated by diamonds). The stages of gravothermal evolution are labeled and described in Sec.~\ref{stages}. The orange curves have larger $\hat{\sigma}$ and thus transition into the SMFP regime earlier.
We indicate the division of stages for each set of curves separately with vertical dashed (dotted) lines for the blue (orange) curves.
{\it (Right)} The quantity $t_{\rm cond} / t_{\rm dyn}$ as a function of $v_c/V_{\rm max}$. The conduction timescale is $t_{\rm cond}=t_{\rm LMFP} + t_{\rm SMFP}$, where $t_{\rm LMFP}=t_{c,0}$ and $t_{\rm SMFP}$ is defined in Eq.~\eqref{t_smfp}. The dynamical timescale $t_{\rm dyn}$ is defined in Eq.~\eqref{eq:t_dyn}. 
}
\label{fig:log_slope_tscale}
\end{figure*}

\subsection{Particle Physics via Conductivity}\label{particlephysics_conductivity}
In this paper, we adopt the analogy of elastic M\o ller scattering in a Yukawa potential~\citep{Girmohanta:2022dog,YangDaneng:2022JCAP...09..077Y}. As in \citetalias{Outmezguine_2022}, we employ a velocity-dependent cross section through the differential cross section of the Born approximation:
\begin{equation}\label{eq:dsig_dOmega}
	\frac{d\sigma}{d\Omega}=\frac{1}{\pi}\frac{\sigma_0 w^4[(3\cos^2\theta+1)v^4+4v^2w^2+4w^4]}{(\sin^2\theta v^4+4v^2w^2+4w^4)^2} ,
\end{equation}
where $w$ is a scale velocity (which is the ratio of the mediator mass to the dark matter mass), $\sigma_0$ is a normalization prefactor, and $\theta$ is the scattering angle. This differential cross section allows us to examine velocity-dependent cross sections, as well as constant cross-sections, as it approaches a constant when the scale velocity $w$ is taken to be very large.
The particle physics model enters the equations only through the heat conductivity, $\kappa$, in Eq.~\eqref{eq:gravothermal}.

\begin{figure*}
\centering
\includegraphics[width=0.96\textwidth]{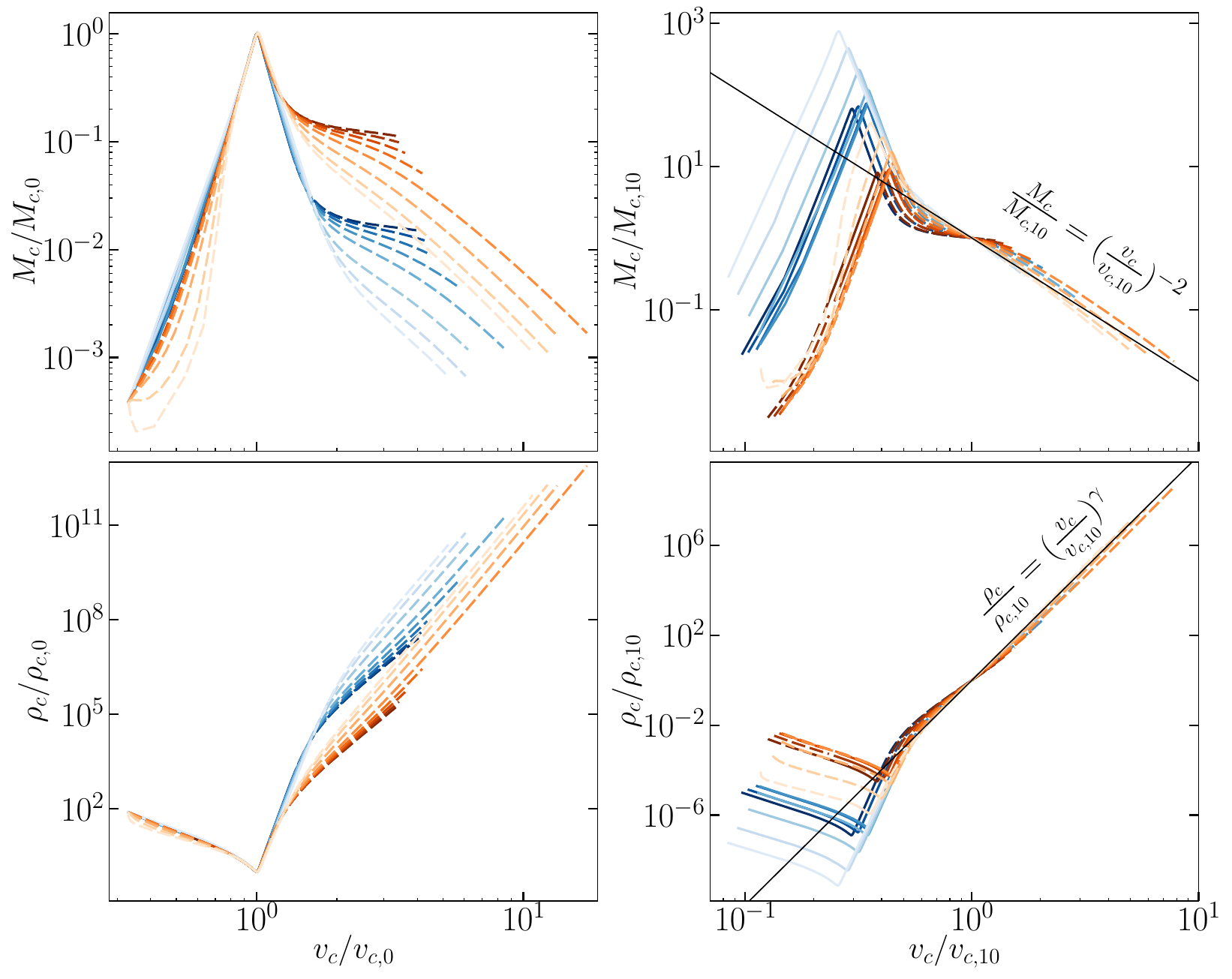}
\caption{ 
{\it(Upper Left)} The evolution of the core mass as a function of the central velocity, normalized by the LMFP scales as in~\citetalias{Outmezguine_2022}, taken at core formation. The dashed lines represent the evolution in the SMFP regime, where for large $v_c$, the scatter in the halos is large. {\it (Upper Right)} Same as panels to the left, except here we normalize with the new scales in the SMFP, $M_{c,10},v_{c,10}$, when the halos have reached the SMFP universal solution at the slope $\gamma=10$. The scatter in the SMFP lines has decreased significantly and shows an approximate universality.
{\it (Lower Left)} Same as figure above, but here it is the central density as a function of velocity. {\it (Lower Right)} Same as figure above, but here it is the central density vs. central velocity, both normalized by new scales $\rho_{c,10},v_{c,10}$. A more detailed explanation of our parameter definitions and notations can be found in Table~\ref{table:param_definition}.
}
\label{fig:mc_rhoc_vc_plots}
\end{figure*}

\begin{figure*}
\centering
\includegraphics[width=0.96\textwidth]{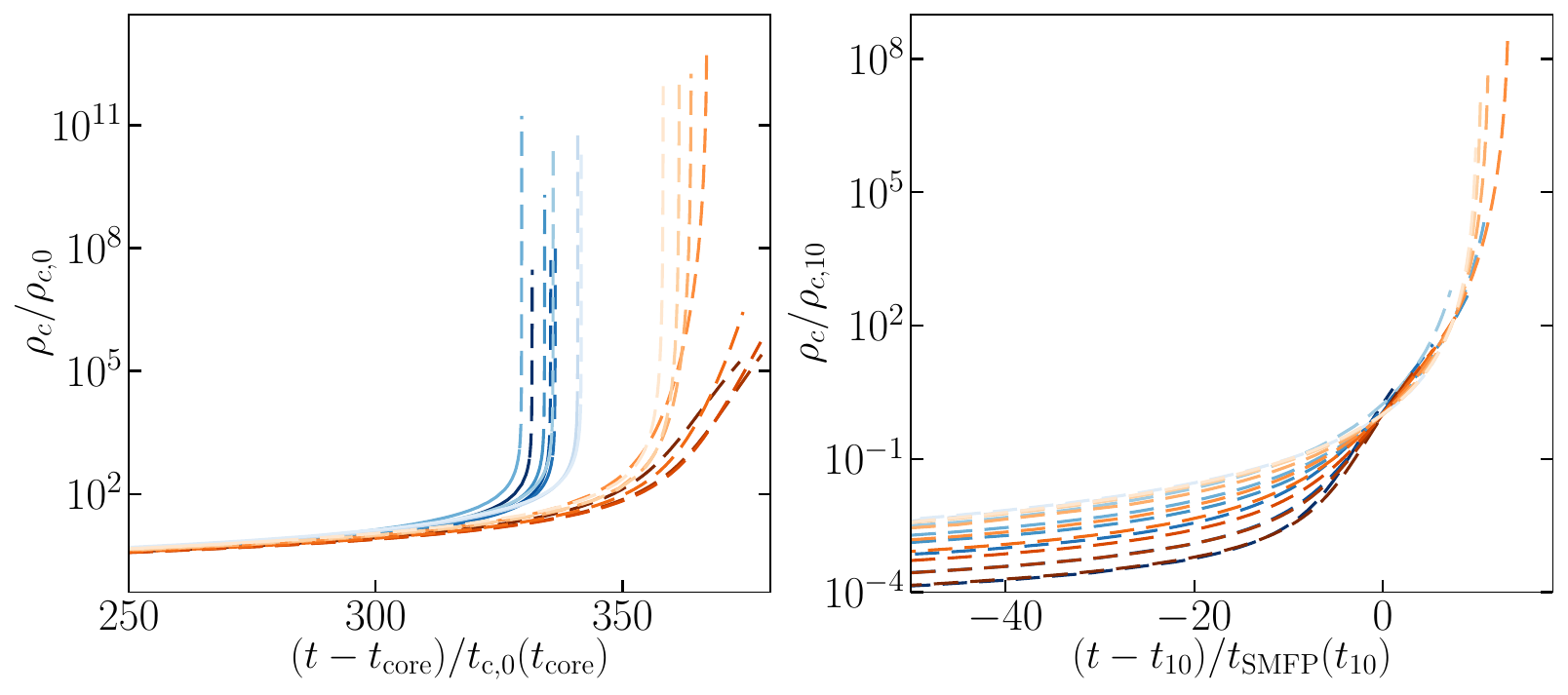}
\caption{ 
{\it(Left)} The evolution of the central density as a function of time, shifted and normalized to the time maximal core is reached as defined in~\citetalias{Outmezguine_2022}. One can see that for different $\sigma_{c,0}$, the collapse times differ. {\it (Right)} Same as the left plot, except the time here is shifted and normalized to the time the halos enter the $\gamma=10$ phase, where the curves line up relatively well as we are already in the self-similar solution. Note that here, the time the curves enter the $\gamma=10$ phase is not analytically determined, thus resulting in the scatter. 
}
\label{fig:rhoc_t_plots}
\end{figure*}

An exact form for the conductivity exists in the SMFP regime, but the LMFP conductivity has only an approximate form and must be calibrated to simulations through a scaling parameter $C$. We follow the notation of \citetalias{Outmezguine_2022} to parameterize the LMFP conductivity as
\begin{equation}\label{eq:kappa_lmfp}
\kappa_{\rm LMFP}=\frac{3aC}{8\pi G}\frac{\sigma_0}{m_{\rm dm}^2}\rho v^3K_5 ,
\end{equation}
where $a=4/\sqrt{\pi}$ and
\begin{equation}\label{eq:K_p}
K_p=\frac{\thb{\sigma_{\rm visc}v_{\rm rel}^p}}{\lim_{w\to\infty}\thb{\sigma_{\rm visc}v_{\rm rel}^p}}.
\end{equation}
Here, $\sigma_{\rm visc}=\int d\sigma \sin^2\theta$ represents the viscosity cross section.
We note that in \citetalias{Outmezguine_2022}, we used $p=3$ in the LMFP conductivity. However, after we released our work,~\cite{Yang_2022} released results of their N-body simulations of SIDM gravothermal collapse which showed that $p=5$ is a better fit. We use their data to find that $C=0.73$ matches best with $p=5$. However, in order to remain consistent with \citetalias{Outmezguine_2022}, we use $C=0.6$, noting that the precise value of $C$ does not impact our main, qualitative results. Moreover, the appropriate value of $C$ for velocity-dependent SIDM requires comparison to simulations, which is beyond the scope of this work.

The SMFP conductivity computed in \citetalias{Outmezguine_2022} is
\begin{equation}\label{eq:kappa_smfp}
\kappa_{\rm SMFP} =\frac{3}{2}\frac{bv}{\sigma_0}\frac{1}{K^{(2)}_{\rm eff}},
\end{equation}
where $b=25\sqrt{\pi}/32$ and
\begin{equation}\label{k_secondorder}
K_{\rm eff}^{(2)}=\frac{28K_5^2+80K_5K_9-64K_7^2}{77K_5-112K_7+80K_9}.
\end{equation}
As discussed in \citetalias{Outmezguine_2022}, we obtained $K_{\rm eff}$ from the Chapman-Enskog expansion \citep{chapman1990mathematical,pitaevskii2012physical} at second order which provides up to a $20\%$ correction to the SMFP heat conductivity. The next order correction results in sub-percent corrections, so we truncate the expansion at second order.
Finally, we interpolate between the LMFP and SMFP regimes as follows:
\begin{equation}\label{eq:kappa_full}
\frac{1}{\kappa}=\frac{1}{\kappa_{\rm LMFP}}+\frac{1}{\kappa_{\rm SMFP}}.
\end{equation}

\subsection{Numerical Methods}
Our numerical procedure closely follows our work in \citetalias{Outmezguine_2022}.
We may recast Eq.~\eqref{eq:gravothermal} to be written in dimensionless form, which depends on only two dimensionless parameters:
\begin{equation}\label{eq:sigma_hat}
    \begin{split}
	\hat{\sigma}^2 &=\frac{aC}{b}K_5\left(\frac{1}{\hat{w}}\right)K_{\rm eff}^{(2)}\left(\frac{1}{\hat{w}}\right)\left(\frac{M_N}{4\pi r_N^2}\frac{\sigma_0}{m_{\rm dm}}\right)^2\\
    &=\frac{aC}{b}K_5\left(\frac{1}{\hat{w}}\right)K_{\rm eff}^{(2)}\left(\frac{1}{\hat{w}}\right)\frac{\rho_{N}}{4\pi G}\left(v_{N}\frac{\sigma_0}{m_{\rm dm}}\right)^2,\;\;\;\;\;\;\; \hat{w}=\frac{w}{v_N},
    \end{split}
\end{equation}
where the scale parameters $\rho_N$, $v_N$, $r_N$, and $M_N$ are related through $GM_N=r_Nv_N^2$ and $M_N=4\pi\rho_N r_N^3$.
Instead of using the parameter $\hat{w}$, we can instead define the quantity $n$ to characterize the velocity-dependent SIDM scattering.
As in \citetalias{Outmezguine_2022}, we use the (positive) log-slope of the function $K$
\begin{equation}\label{eq:n_func}
	n=-\frac{d\log K}{d\log v_N},
\end{equation}
where $K=K_5$ in the LMFP regime and $K=K_{\rm eff}^{(2)}$ in the SMFP regime.

We employ our gravothermal code used in \citetalias{Outmezguine_2022} to evolve velocity-dependent SIDM halos according to the dimensionless form of Eq.~\eqref{eq:gravothermal}.
We assume halos initially have an NFW profile, and from \citetalias{Outmezguine_2022}, the NFW parameters are related to the central parameters at core formation as $\rho_{c,0}\simeq2.4\rho_s,\,v_{c,0}\simeq0.64V_{\rm max}$ (see Table~\ref{table:param_definition} for a description of our notation and definitions).
We evolve 16 halos with the parameters listed in Table~\ref{table:run_params}.
For the purpose of maintaining numerical stability, we approximate fitted forms for Eq.~\eqref{eq:K_p}, which we provide in Appendix~\ref{appendix_approx}.

We define the size of the core of a halo $r_{\rm core}$ via $\rho(r_{\rm core})=\rho_c/2$.
As the halo evolves, different regions of the halo may transition from the LMFP regime (for which $\kappa_{\rm LMFP} < \kappa_{\rm SMFP}$) to the SMFP regime (for which $\kappa_{\rm LMFP} > \kappa_{\rm SMFP}$).
We use the condition
\begin{equation}\label{kappa_ratio}
	\left. \frac{\kappa_{\rm SMFP}}{\kappa_{\rm LMFP}} \right|_{r=r_{\rm core}}=1
\end{equation}
to define the time when the core fully transitions from the LMFP to SMFP regime, and we refer to this time as the LS transition.

\section{Long and Short Mean Free Path Evolution}\label{stages}

We can delineate a halo undergoing gravothermal collapse into two regions: the outer halo and the central core. Gravothermal evolution leads to core collapse: the core contracts and heats up, leading to a high density and temperature, during which time dark matter self-interactions transport heat and mass to the outer halo. Meanwhile, the outer halo that surrounds the core remains relatively dilute and acts as a heat sink. The thermal evolution timescale of the outer halo is much longer than that of the core, so the outer halo changes temperature very slowly, while the core temperature increases rapidly.
The outer halo remains in the LMFP regime, while the core can be in the LMFP, the SMFP, or an intermediate regime. The evolution in the two regimes differs substantially. In \citetalias{Outmezguine_2022}, we explored the LMFP regime in detail. In this paper, we focus on the SMFP regime. We provide a brief overview below of the two regimes.

In the LMFP regime, dark matter particles are sufficiently dilute such that particles can make many orbits before scattering. 
While the outer halo is always in the LMFP regime, for the core 
of a halo, this is not always the case.
Given a large enough cross section, the core of a halo can begin in the SMFP regime. However at core formation, as we studied in depth in \citetalias{Outmezguine_2022}, the cores of halos evolve in the LMFP regime, and the LMFP solution represents the bulk of the time an isolated halo spends in its evolution. The self-scattering enables dark matter particles to transfer heat and mass between the inner and outer regions of the halo. Initially, heat is transferred inward from the hotter outer halo to the colder inner core, causing the core to expand until it reaches a maximum, or ``maximal core.'' The core then transfers heat to the outer halo and shrinks, becoming denser and hotter. This runaway process inevitably drives the core into the SMFP regime and into the core collapse phase.
The LMFP and SMFP regimes experience distinct evolution paths. This is illustrated in the curves of Figs.~\ref{fig:log_slope_tscale} (left panel) and \ref{fig:mc_rhoc_vc_plots} (all panels) where the dashed portions of the curves denote when the core 
is in the SMFP regime, whereas the solid portions indicate the LMFP regime. We delineate the LS transition as the point when the conductivities are equal, as in Eq.~\eqref{kappa_ratio}.
Notably, the orange curves transition from SMFP (dashed) at low $v_c$ values to LMFP (solid) and return to SMFP (dashed) at high $v_c$ values. In contrast, the blue curves begin in the LMFP regime (solid) and switch to the SMFP regime (dashed) as $v_c$ values increase. Thus, halos with substantial cross sections begin their evolution with their cores in the SMFP regime. Nonetheless, we assume NFW profiles as an initial condition, as we did in~\citetalias{Outmezguine_2022}, noting that halo core formation causes the core to transition to the LMFP regime early in its evolution. It would be interesting to test other profiles with different inner slopes, for example, a roughly constant core. We know that profiles that are truncated in the outer regions due to tidal effects tend to speed up the onset of core collapse~\citep{Sameie_2018MNRAS.479..359S,Kahlhoefer:2019,Nishikawa_2020,Correa:2021}, and thus it would be insightful to see how profiles with inner slopes that differ from $\sim r^{-1}$ may affect the evolution. We leave this interesting question to a future study.

\begin{figure*}
\centering
\includegraphics[width=0.98\textwidth]
{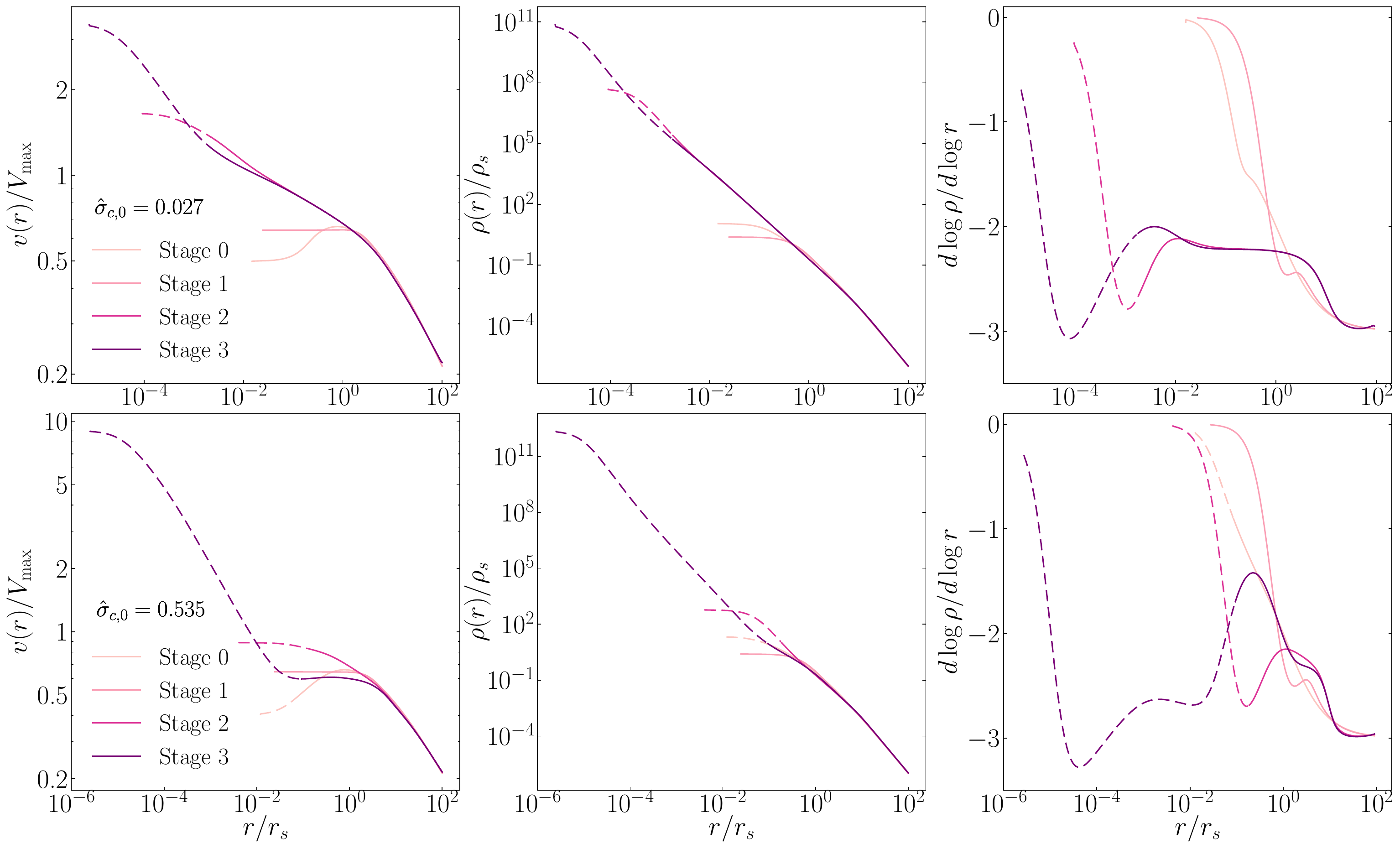}\;\;\;\;
\caption{ 
{\it(Left)} The velocity dispersion as a function of radius for Stages 0, 1, 2, and 3. The upper row is for a small $\hat{\sigma}$ with $n=3.7$ corresponding to run 8 in Table~\ref{table:run_params}, and the lower row is for large $\hat{\sigma}$ with $n = 1$ corresponding to run 13 in Table~\ref{table:run_params}. {\it(Middle)} Density as a function of radius. {\it (Right)} Slope of the density profile with respect to radius.
}
\label{fig:profiles}
\end{figure*}

In the SMFP regime, the core shrinks and drastically increases in density.
In this regime, particles are so dense that collisions are very frequent. The dense core as a result hampers heat and mass transfer from the core to the outer halo~\citep{Balberg:2002ue,Agrawal_2017JCAP...05..022A,Essig:2018pzq}. After the shielding begins, the evolution paths of the outer and core become distinct, and the core can be treated approximately independently of the outer halo. 
Shortly after this transition, the core falls into a phase of constant thermal energy, where the log-slope of the central density with respect to central velocity dispersion becomes constant in time, approaching a slope of $\gamma=10$; see Fig.~\ref{fig:log_slope_tscale}. Here, the core heats up very quickly and the density shoots up by several orders of magnitude, which can be seen as a steady rise of the central density with respect to the central velocity in Fig.~\ref{fig:mc_rhoc_vc_plots} (lower panels), as well as both panels of Fig.~\ref{fig:rhoc_t_plots} where the central density shoots up during collapse. Looking at both Figs.~\ref{fig:mc_rhoc_vc_plots} (lower panels) and~\ref{fig:rhoc_t_plots}, it is also evident that the timescale at which the LMFP and SMFP evolution proceeds is drastically different: while the bulk of the evolution resides in the LMFP, the SMFP evolution occurs very rapidly. 

In the left panel of Fig.~\ref{fig:log_slope_tscale}, we label each stage in the gravothermal evolution. We call the quantity $d\log\rho_c/d\log v_c$ as $\gamma$, given by 
\begin{equation}\label{gamma_eq}
    \frac{d\log\rho_c}{d\log v_c}=\gamma
\end{equation}
where $\rho_c\propto v_c^\gamma$ with $\gamma\simeq10$. 
Fig.~\ref{fig:profiles} shows the density profiles for each of these stages.
The outer halo remains largely unchanged throughout the evolution, but the core changes significantly. 
The stages are as follows: 
\begin{itemize}[leftmargin=*, labelindent=1pt, itemsep=.2cm, parsep=0pt]
    \item \textbf{Stage 0}: The Core Expansion Phase. During this initial phase, the halo's core undergoes an expansion process. While the core size increases, it experiences adiabatic heating and a concurrent gradual decrease in density.
    \item \textbf{Stage 1}: The Core Contraction Phase. This phase commences once the core reaches its maximal size and minimal central density, initiating a period of contraction. During this phase, the core decreases in size and increases in density. This marks the onset of the LMFP evolution phase, characterized by self-similarity and approximate LMFP universality, explored extensively in \citetalias{Outmezguine_2022}.
    \item \textbf{Stage 2}: The Unorganizable Phase. This phase signifies a departure from the self-similarity and universality found in the earlier stages. Here, the halo does not conform to any universally applicable solution. Despite this, we have developed semi-analytical formulae that effectively map the values of macroscopic parameters at the end of Stage 1 to those at the beginning of Stage 3. Detailed discussions on these formulae are provided in \S\ref{SMFP_universality}.
    \item \textbf{Stage 3}: The Constant Thermal Energy Phase. This phase represents the region beyond the $\gamma=10$ transition where we identify a new SMFP universality. As depicted in the left panel of Fig.~\ref{fig:log_slope_tscale}, all halos tend towards the slope $\gamma=10$ during this stage, allowing us to fully characterize the evolution during Stage 3.
\end{itemize}

\section{SMFP Approximate Universality}
\label{SMFP_universality}

In the LMFP regime, the self-similar solution is most clearly seen by scaling the central (i.e., at the halo center) quantities $\rho_c,\, r_{\rm core},$  and $v_c$
by their value at core formation, which we refer to as the instant of maximal core in \citetalias{Outmezguine_2022}. It represents the time at which the central density is at a minimum and the size of the core is at a maximum (see the left panels of Fig.~\ref{fig:mc_rhoc_vc_plots}). The approximate universality, however, breaks once the core enters the SMFP regime, as evident in the left columns of Figs.~\ref{fig:mc_rhoc_vc_plots} and~\ref{fig:rhoc_t_plots}. As discussed in \citetalias{Outmezguine_2022}, the breakdown of universality at the LS transition is governed by the size of $\hat{\sigma}$. 

In this section, we present a new approximate universality in the deep SMFP regime. This universality is evident upon scaling the central quantities by their respective values at the onset of Stage 3, as seen in the right column of Fig.~\ref{fig:mc_rhoc_vc_plots} and~\ref{fig:rhoc_t_plots}.
The onset of Stage 3 of the evolution is not as clear-cut as the LS transition and cannot be computed analytically; however, as previously stated, we approximate the point of the $\gamma=10$ transition by taking the minimum of the curves in the SMFP regime in the left panel of Fig.~\ref{fig:log_slope_tscale}. The evolution of the newly scaled quantities in the SMFP regime all align, as seen in the right panels of Figs.~\ref{fig:mc_rhoc_vc_plots} and~\ref{fig:rhoc_t_plots}.

\subsection{Analytic description of the Core Structure}
\label{core_structure}
We now show some analytical properties of the halo cores during Stage 3 in the deep SMFP regime. Detailed derivations are presented in Appendix \ref{appendix_derivations}.

\begin{figure}
\centering
\includegraphics[width=0.48\textwidth]{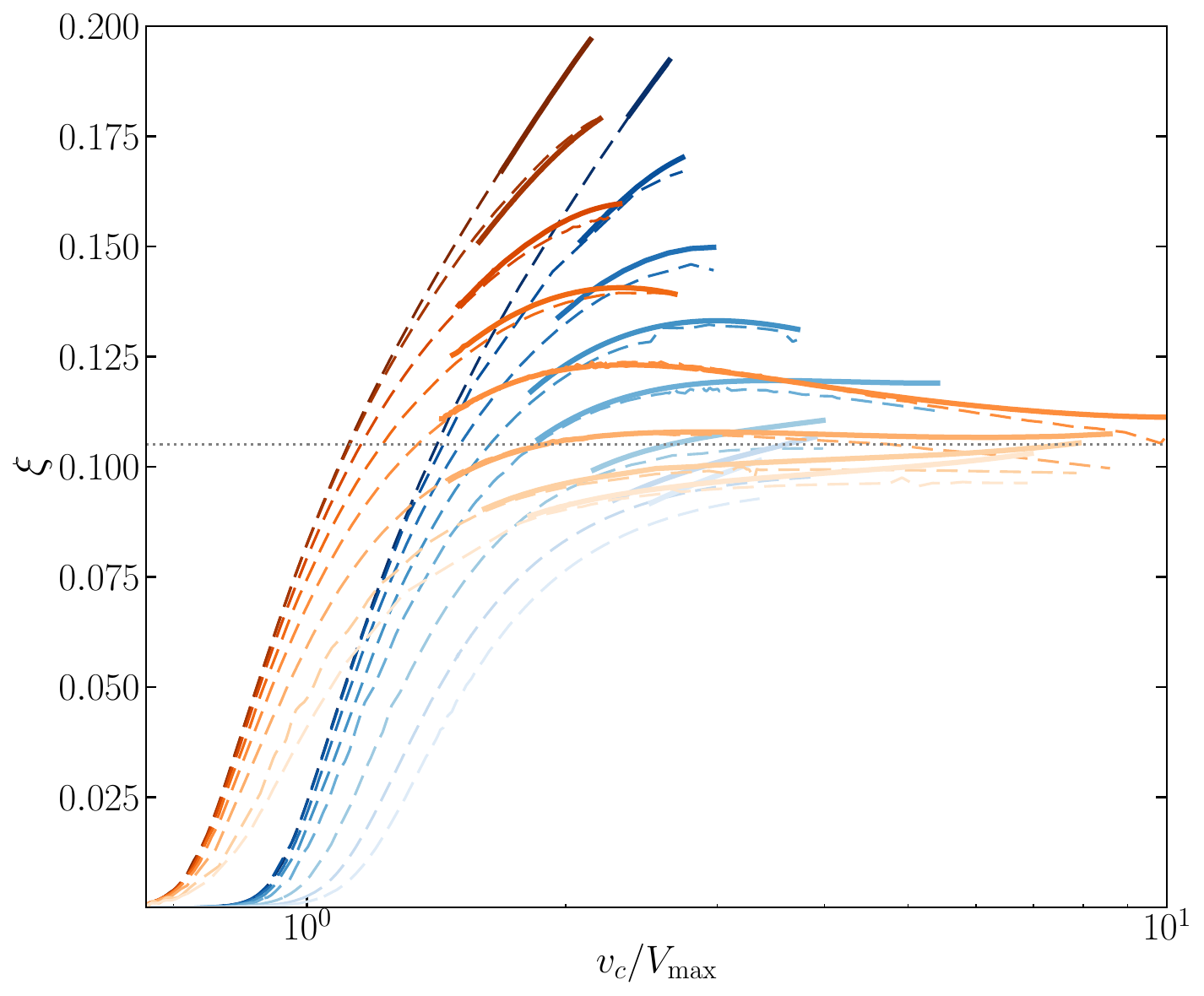}
\caption{ $\xi$ as a function of $v_c/V_{\rm max}$. The dashed curves are obtained using Eq.~\eqref{eq:d_log_Sc_dt} for $\xi$, and the solid ones obtained by using Eq.~\eqref{eq:xi_main_r} for $\xi$. The dotted line indicates the line to which the curves tend to, which is $\xi\simeq0.11$.
}
\label{fig:xi_plots}
\end{figure}

We start by Taylor expanding the halo density and velocity dispersion profiles around the halo center:
\begin{equation}\label{rho_v_small_r}
    \begin{split}
	\rho &= \rho_c\left(1-x^2 + {\cal O}(x^4)\right), \\ v^2 &= v_c^2\left(1-\xi x^2 +\xi\beta x^4 + {\cal O}(x^6)\right),
    \end{split}
\end{equation}
where $x=r/r_c$, and we have implicitly defined a core radius $r_c$ as
\begin{equation}\label{core_radius}
    r_c^{-2} =-\lim_{r\to0}\left(\frac{\partial\log\rho}{\partial r^2}\right)_t\,.
\end{equation}
We note that $\xi$ can be calculated from our numerical results:
\begin{equation}\label{eq:xi_main_r}
    \xi=\lim_{r\to 0}\left(\frac{\partial \log v^2}{\partial \log \rho}\right)_t .
\end{equation}
To leading order in $x$, mass conservation gives $M\simeq M_c x^3$, with $M_c=4\pi\rho_cr_c^3/3$.
Using the above expansion to the lowest order in the Jeans equation in Eq.~\eqref{eq:gravothermal}, we find that the central quantities are related through
\begin{equation}\label{v_core}
    v_c^2 = \frac{GM_c}{2 r_c (1+\xi)}.
\end{equation}
As discussed earlier, we find that during Stage 3, $\rho_c\propto v_c^\gamma$ with $\gamma\simeq 10$.
From the above relations, we see that $M_c\propto v_c^{3-\gamma/2}\simeq v_c^{-2}$, or
\begin{equation}\label{m10_v10_scaling}
    \frac{M_c}{M_{10}}=\left(\frac{v_c^2}{v_{10}^2}\right)^{-1}.
\end{equation}
This scaling is nicely demonstrated in the bottom right panel of Fig.~\ref{fig:mc_rhoc_vc_plots}. It is important to point out that $\frac{d\log M_c}{d\log v_c^2}$ is not exactly $-1$, as a value of $-1$ indicates no core evolution~\citep{Balberg:2002ue}. But it gets close to $-1$ in our runs and so we use this approximation. 
We define the SMFP timescale as
\begin{equation}\label{t_smfp}
    t_{\rm SMFP}=\frac{\sigma_0 K_{\mathrm{eff}}^{(2)}\left(v_c\right) v_c}{2 \pi b G m_{\mathrm{dm}}},
\end{equation}
and find in Appendix \ref{appendix_derivations} that 
\begin{equation}\label{eq:xi_main_t}
    \lim_{r\to0}\left(\frac{\partial \log(v^3/\rho)}{\partial t}\right)_r=-\frac{3}{t_{\rm SMFP}}\frac{\xi}{1+\xi}.
\end{equation}
Note that through Eqs.~\eqref{eq:xi_main_r} and~\eqref{eq:xi_main_t}, we obtain two different derivations of $\xi$: one through a temporal derivative and the other through spatial ones. The agreement between the two different derivations is demonstrated in Fig.~\ref{fig:xi_plots}.

It is not clear exactly why this approximate universality exists. Unlike the LMFP universality we found in \citetalias{Outmezguine_2022}, the SMFP universality does not have a simple analytic derivation, and we can only approximate the $\gamma=10$ transition numerically. But once this approximation is done, the universality does appear upon rescaling, and we produce semi-analytic expressions for these parameters using the fits in Eq.~\eqref{nls_fits}.

Although the particle physics dependence (the velocity dependence of the cross-section) can be scaled out in Stage 1 (the LMFP) and Stage 3 ($\gamma=10$ transition), Stage 2 (the drop from the LS transition to the $\gamma=10$ transition in the left panel of Fig.~\ref{fig:log_slope_tscale}) seems to have a non-trivial particle physics dependence. Accounting for this dependence is crucial to getting some analytic handle on the $\gamma=10$ transition, which can be related to the LS transition, as shown in \S\ref{ls_to_gamma10}.

\subsection{SMFP timescale and validity of the gravothermal equations}

In the LMFP regime, changes in the evolution occur on the scattering timescale given by
\begin{equation}\label{eq:t_LMFP}
	t_{N}= \frac{2}{3aC}\left(\rho_{N}\frac{\sigma_0}{m_{\rm dm}}v_{N} K_5\left(\frac{v_{N}}{w}\right)\right)^{-1}
\end{equation}
for which the core parameters are taken at core formation in the LMFP (see Table~\ref{table:param_definition} for a full explanation of our definitions and notations at core formation, the LS transition, and $\gamma=10$ transition). In the SMFP regime, the timescale is given in Eq.~\eqref{t_smfp}
where, as in Eq.~\eqref{eq:sigma_hat}, the core parameters can be taken at the LS or $\gamma=10$ transitions depending on the stage of evolution.
For the gravothermal equations to be valid, the relevant timescale must always be longer than the dynamical time
\begin{equation}\label{eq:t_dyn}
	t_{\rm dyn}=\frac{1}{\sqrt{4\pi G\rho_N}}.
\end{equation}
Note that $t_{\tx{SMFP}}\propto t_\tx{dyn}^2/t_{c,0}$, where $t_{c,0}$ is Eq.~\eqref{eq:t_LMFP} taken at core formation. One may be concerned that the condition $t_{\rm SMFP} > t_{\rm dyn}$ is not fulfilled, because changes in the SMFP occur very rapidly. In the right panel of Fig.~\ref{fig:log_slope_tscale}, note that $t_{\rm cond}/t_{\rm dyn}>1$, and while it gets close to 1, it does so in the LMFP regime, which has been investigated in simulations that show the gravothermal equations are valid; see for example \cite{YangDaneng:2022JCAP...09..077Y,Yang_2022}. Once the halos are in the SMFP regime (dashed lines), $t_{\rm cond}/t_{\rm dyn}>1$ is always true. As long as the timescale of SMFP evolution is larger than the dynamical timescale, changes due to the evolution take longer than the dynamical time of the halo, allowing the halo to maintain hydrostatic equilibrium. Thus, the gravothermal equations remain valid.

\section{Core mass at the Relativistic Instability}\label{black_hole_estimates}
Previous works have made estimates of the core mass at the relativistic instability~\citep{Balberg:2002ue,Koda:2011yb,Nishikawa_2020,Meshveliani_2022}. However, lacking the properties of the halo in Stage 3 that we have found has led to an overestimation of this mass. The scaling that has been used to extrapolate to high temperatures is $\frac{d\log M_c}{d\log v_c^2} = -0.85$~\citep{Balberg:2002ue,Koda:2011yb,Meshveliani_2022}. Extrapolating to the relativistic instability velocity, which is estimated to be $v_c \simeq c/3$~\citep{Feng_2022JCAP...05..036F}, this results in a substantial overestimation of the core mass. Examining the upper panels of Fig.~\ref{fig:mc_rhoc_vc_plots}, it becomes evident that the log slopes of the curves initially flatten to approximately $-0.85$ shortly after entering the SMFP regime (marked by dashed lines). This is followed by a steepening of the slopes, eventually converging to a log-slope of about $-1$. As we explained in \S\ref{core_structure}, the slope cannot be exactly $-1$, but it is very close to it and steeper than $-0.85$.

In this section, we present a new, more rigorous, prescription to connect the core parameters at the LS transition to those in Stage 3 at the $\gamma=10$ transition. 

\begin{figure*}
\centering
\includegraphics[width=0.488\textwidth]{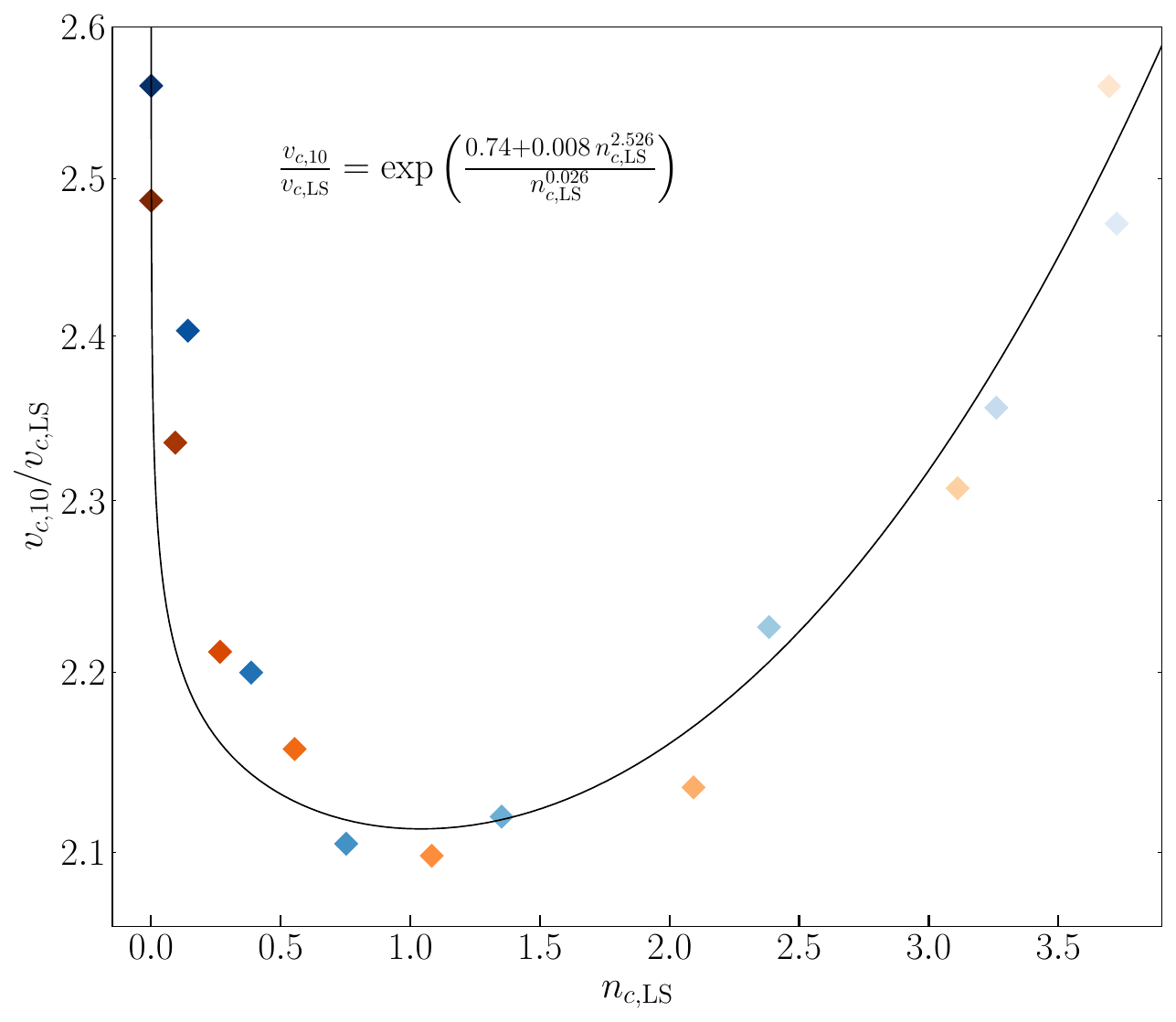}\;\includegraphics[width=0.49\textwidth]{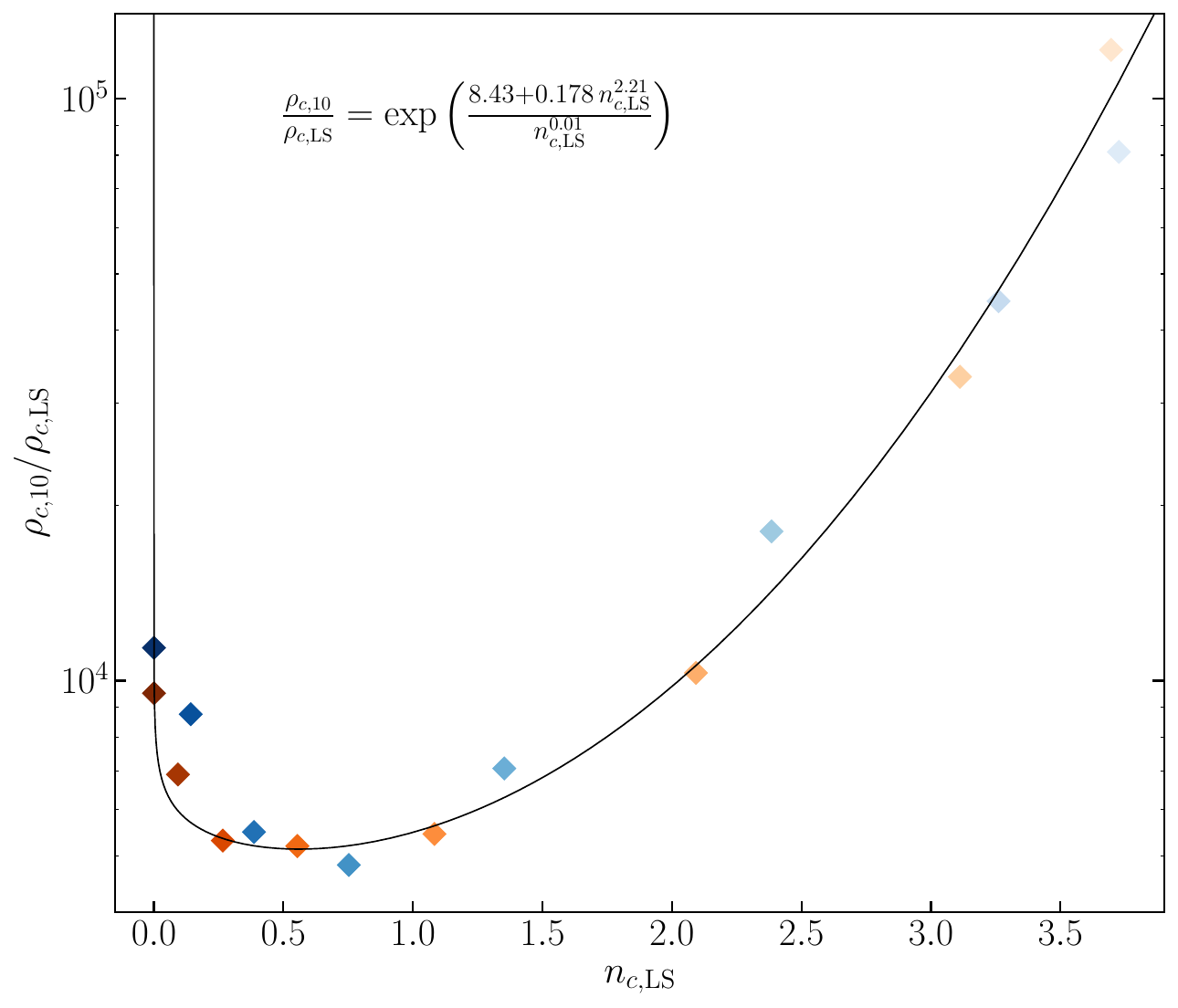}
\caption{ 
{\it(Left)} The $v_{c,10}/v_{c,\rm LS}$ to $n_{c,\rm LS}$ relation plotted as a black curve, fitted to our numerical results (diamonds). {\it (Right)} Same as the left, but for the $\rho_{c,10}/\rho_{c,\rm LS}$ to $n_{c,\rm LS}$ relation.
}
\label{fig:v10_vls_rho10_rhols}
\end{figure*}

\begin{figure*}
\centering
\includegraphics[width=0.49\textwidth]{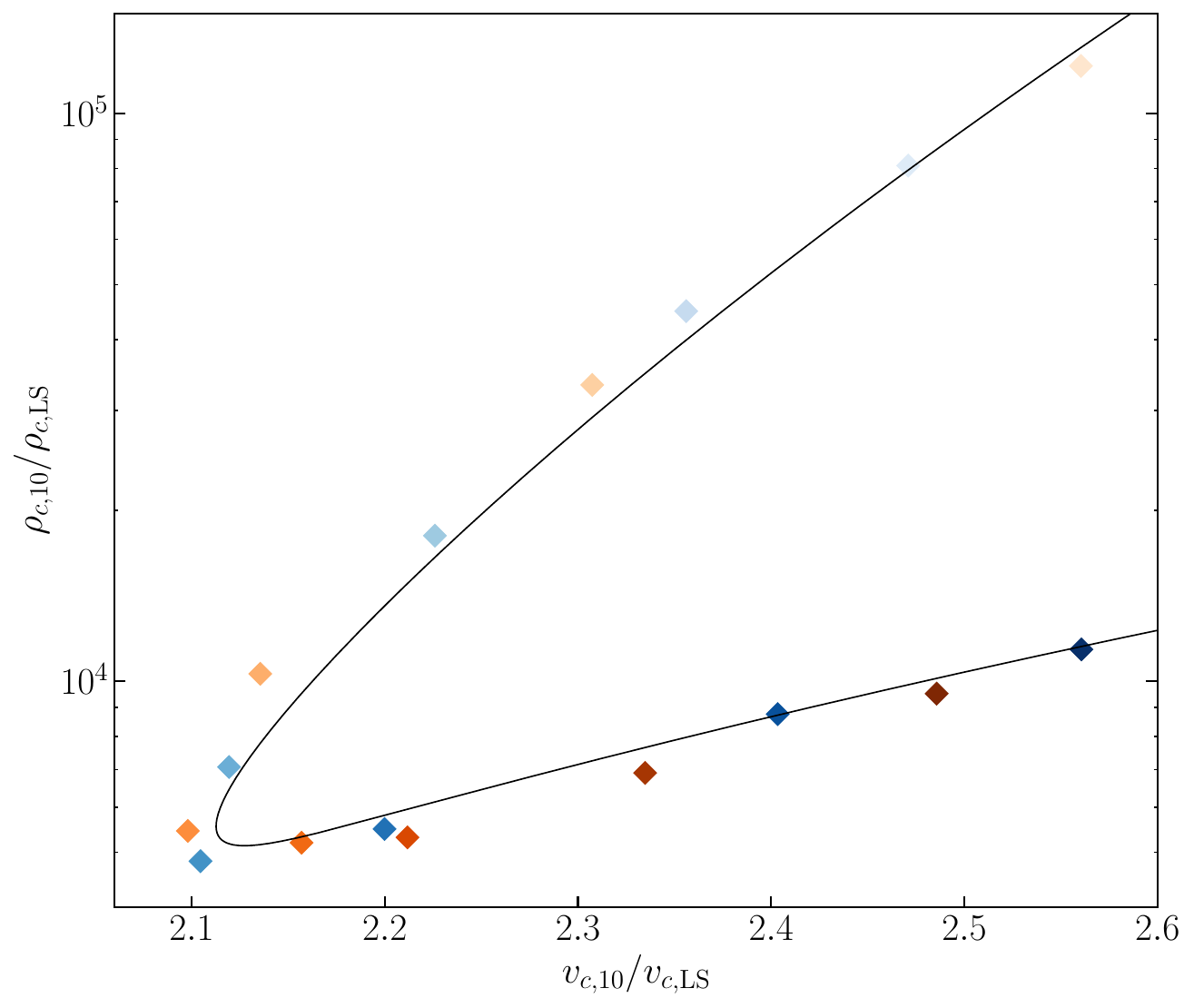}\;\includegraphics[width=0.483\textwidth]{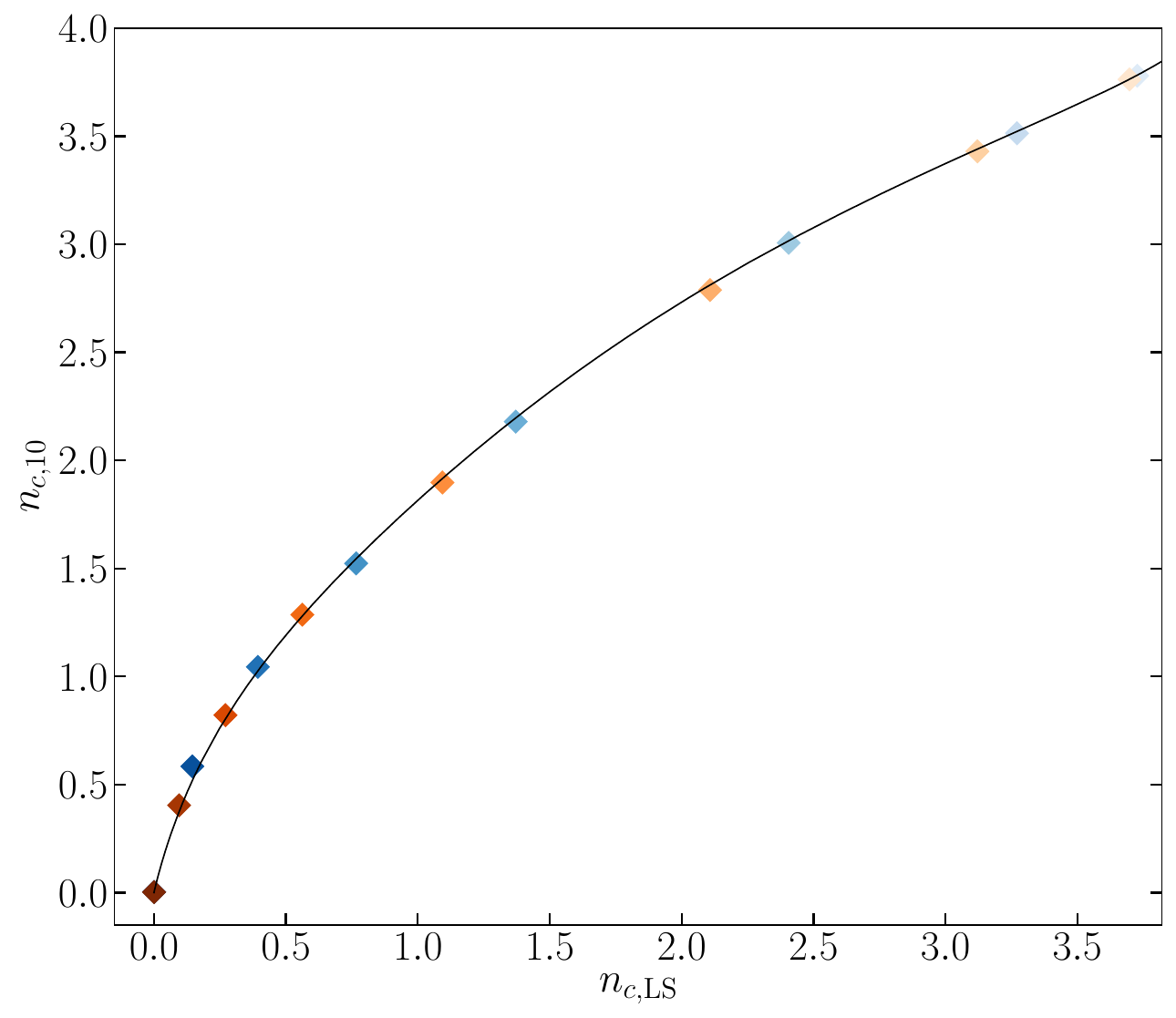}
\caption{ 
{\it(Left)} $\rho_{c,10}/\rho_{c,\rm LS}$ vs $v_{c,10}/v_{c,\rm LS}$ shown as a black curve fitted using the fits in Figs.~\ref{fig:v10_vls_rho10_rhols}, plotted over our numerical runs (diamonds).  {\it (Right)} Plot of the $n_{c,10}$ to $n_{c,\rm LS}$ relation, fitted using the same relation as in the left panel of Fig.~\ref{fig:v10_vls_rho10_rhols}.
}
\label{fig:rhoratio_vratio_nfits}
\end{figure*}

\begin{figure*}
\centering
\includegraphics[width=0.33\textwidth]{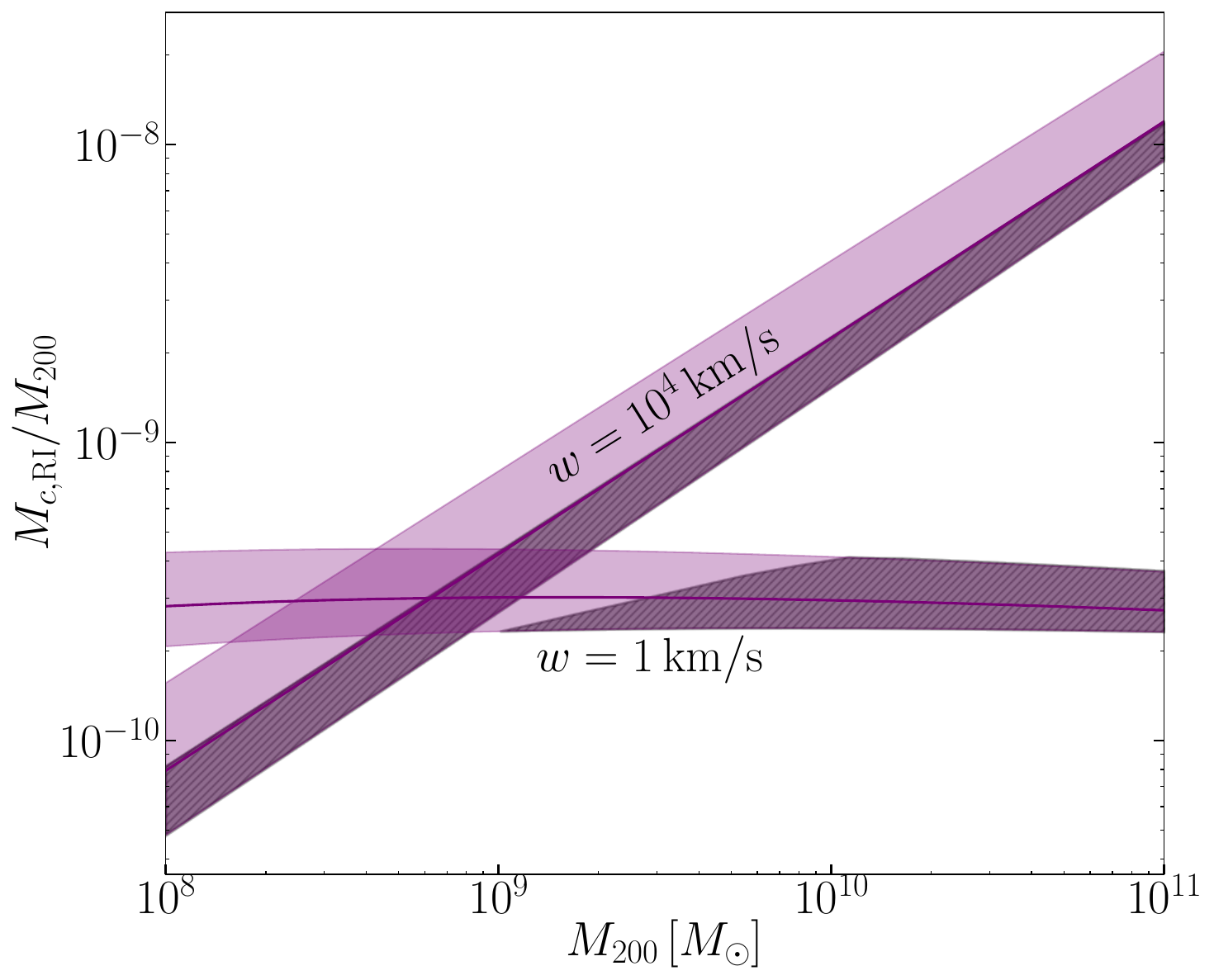}\includegraphics[width=0.33\textwidth]{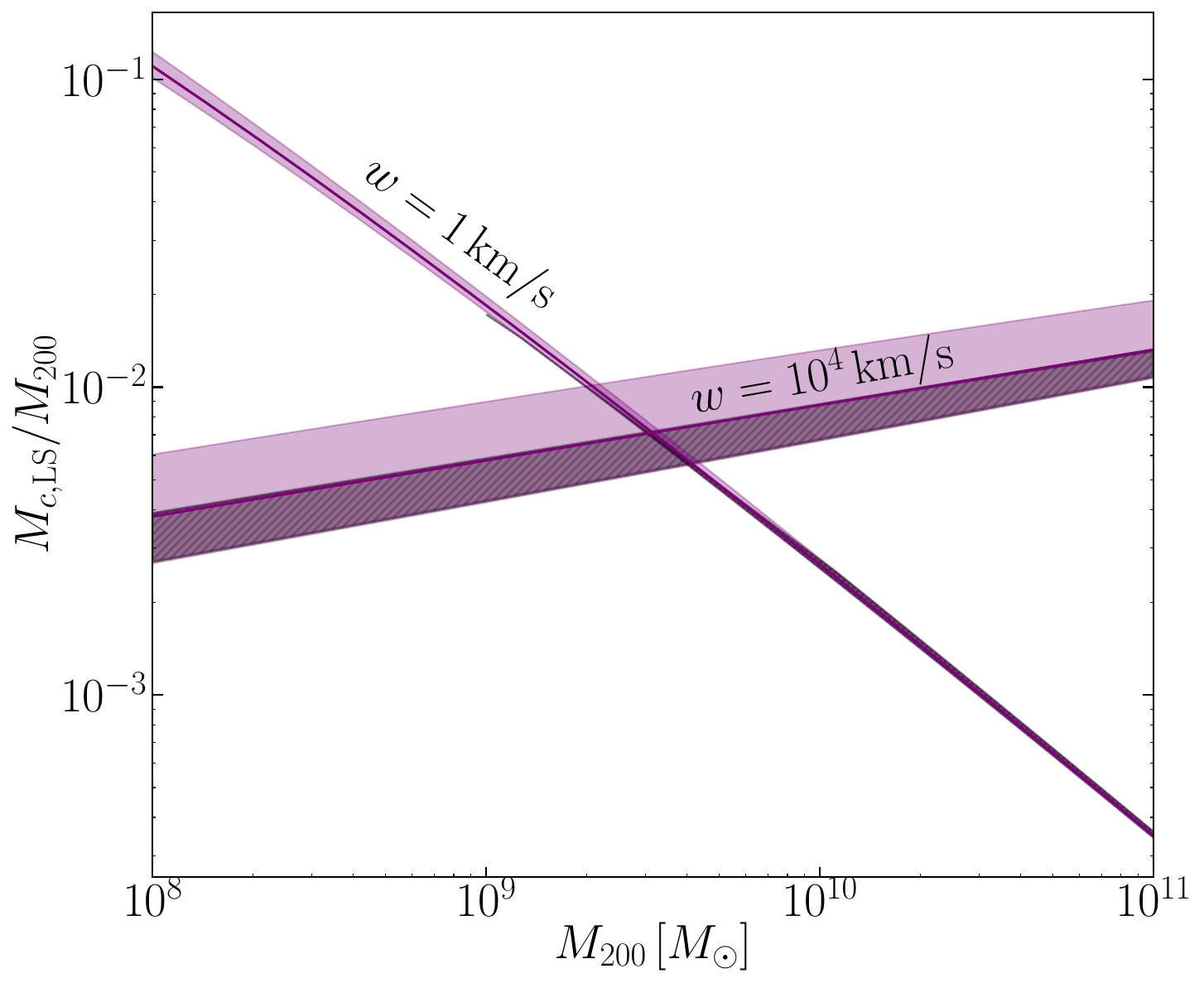}\includegraphics[width=0.325\textwidth]{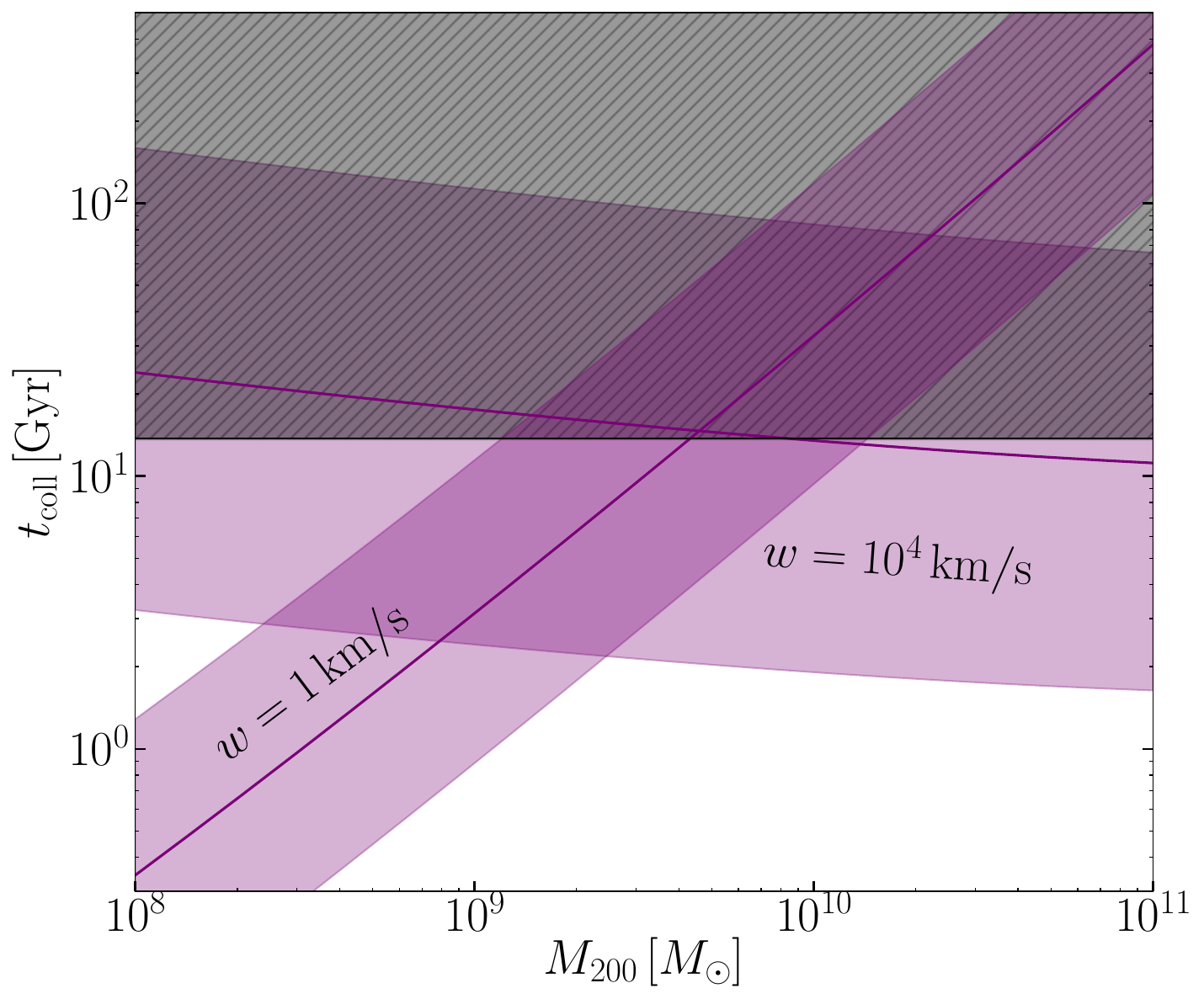}

\caption{ 
{\it(Left)} The ratio of $M_{c,\rm RI}/M_{200}$ with respect to the halo mass $M_{200}$ for a cross section of $\sigma_0K_p=100\,\rm{cm^2/g}$ at $v=20\,\rm km/s$ with $w=1\,\rm{km/s}$ and $w=10^4\,\rm{km/s}$ (labeled). The band for the velocity independent model overtakes the velocity dependent one for masses larger than $10^{9}\,M_\odot$. The gray hatched shaded regions are those that do not collapse within the age of the Universe (see right panel). {\it(Middle)} Same as the left panel, but for the core mass at the LS transition. {\it(Right)} Collapse time as a function of the halo mass for the same models as in the left panel, with the gray hatched shaded region above the line at ($13.7\,\rm Gyr$) indicating collapse times longer than the age of the universe. The bands for the cases of $w=10^4\,\rm{km/s}$ and $w=1\,\rm{km/s}$ are labeled in the figure. The purple shaded bands in all the figures cover a concentration of $\pm 0.3$ dex around the median, represented by the dark purple line. In the first two figures, the upper (lower) bound of the bands correspond to higher (lower) concentration, while in the rightmost figure, the upper (lower) bound represents the lower (higher) concentration, indicating that higher concentration halos collapse faster.}
\label{fig:mbh_m200_relation}
\end{figure*}

\subsection{Relating the LS transition to the Stage 3 Parameters}\label{ls_to_gamma10}

As elaborated in Section 4, the onset of the newly identified universal phase—referred to as Stage 3—commences at the $\gamma=10$ transition, for which we lack exact analytical predictions. In this section, we introduce empirical formulae that allow for the estimation of key halo core characteristics, starting from the core parameters at the LS transition. Given that these LS parameters can be analytically derived from the initial core characteristics, we essentially offer a semi-analytical framework to predict the core parameters at the $\gamma=10$ transition based on the initial halo conditions and particle physics parameters. Using the approximate universality described in Section 4, this, in principle, allows for a semi-analytical halo evolution model throughout the different phases.

The relation between the LS and $\gamma=10$ parameters is shown in Fig. \ref{fig:v10_vls_rho10_rhols}, alongside the fitted lines given below:

\begin{equation}\label{nls_fits}
    \begin{split}
        \frac{v_{c,10}}{v_{c,\tx LS}}&=\exp\left(\frac{0.74+0.008n_{c,\tx LS}^{2.5}}{n_{c,\tx LS}^{0.03}}\right) \\
        \frac{\rho_{c,10}}{\rho_{c,\tx LS}}&=\exp\left(\frac{8.43+0.18n_{c,\tx LS}^{2.21}}{n_{c,\tx LS}^{0.013}}\right)
    \end{split}
\end{equation}
The fitted lines closely align with the behavior of the numerically
calculated points. The above fits also introduce an implicit relation
between $v_{c,10}/v_{c,\rm LS}$ and $\rho_{c,10}/\rho_{c,\rm LS}$, this implicit relation is the line shown on the left panel of Fig.~\ref{fig:rhoratio_vratio_nfits}. This line follows the numerical calculations quite accurately.  Per the definition in Eq.~\eqref{eq:n_func}, the log-slope, $n$, is a function of the
central velocity dispersion, namely $n=n(v_c/w)$. Denoting the inverse relation by $v_c/w = f(n)$ we can express $v_{c,10}/v_{c,\rm LS} = f(n_{c,10})/f(n_{c,\rm LS})$, solving for $f(n_{c,\rm LS})$, and noting that $n[f(x)]=x$, we find that:
\begin{equation}\label{eq:n_ls}
    n_{c,\rm LS} = n\left[\frac{v_{c,\rm LS}}{v_{c,10}} f(n_{c,10})\right]\,.
\end{equation}
The resulting line on the right panel of Fig.~\ref{fig:rhoratio_vratio_nfits} validates this approach, as it closely matches our numerical calculations.

To complete our semi-empirical model, we need to provide a means for finding $n_{c,\rm LS}$, $v_{c,\rm LS}$ and $\rho_{c,\rm LS}$ from the initial halo parameters. In \citetalias{Outmezguine_2022} we derived the scaling laws
\begin{equation}\label{eq:v_rho_LS}
    \frac{v_{c,\rm LS}}{v_{c,0}}\simeq\hat{\sigma}_{c,0}^{-1/\delta}\,,\,\frac{\rho_{c,\rm LS}}{\rho_{c,0}}\simeq\left(\frac{v_{c,\rm LS}}{v_{c,0}}\right)^\frac{2\alpha}{\alpha-2}\,,\,
    \frac{M_{c,\rm LS}}{M_{c,0}}\simeq\left(\frac{v_{c,\rm LS}}{v_{c,0}}\right)^{\frac{6-2\alpha}{2-\alpha}}.
\end{equation}
where $\delta\simeq1-n_{c,0}+\alpha/(\alpha-2)$, and $\alpha\sim2.2$ up to small but important scatter which is detailed in \citetalias{Outmezguine_2022}. With this at hand, we simply calculate $n_{c,\rm LS}$ by using Eq.~\eqref{eq:n_ls}. Having these relations allows us to have a semi-analytic handle on the $\gamma=10$ region, allowing one to obtain the parameters in the $\gamma=10$ region by simply having the initial halo and cross-section model. Eq.~\eqref{eq:v_rho_LS} relies on the assumption that core formation occurs in the LMFP, which is satisfied when $\kappa_{\rm SMFP}\gg\kappa_{\rm LMFP}$ at the time of core formation, $t_{\rm core}$. This implies that $\hat{\sigma}\ll1$ [cf. Eqs.~\eqref{eq:kappa_lmfp}, \eqref{eq:kappa_smfp}, and~\eqref{eq:sigma_hat}].
However, our numerical results demonstrate the effectiveness of Eq.~\eqref{eq:v_rho_LS} even in scenarios where $\hat{\sigma}\sim1$, highlighting its applicability beyond its initial theoretical assumptions.

\subsection{Obtaining the core mass at the relativistic instability from halo properties}\label{analytic_core_mass_ri}
Equipped with the results of the past section, we can derive updated predictions for the core mass at the onset of the relativistic instability. Before doing so, it is crucial to emphasize that the core mass at the point of relativistic instability, $M_{c,\rm RI}$, should be considered a lower limit for the mass that will ultimately collapse to form a black hole. We conjecture that the upper bound for black hole growth due to processes such as accretion is the mass within the region of the halo that is in the SMFP, and can be approximated as the core mass of the halo at the LS transition, $M_{c,\rm LS}$ (see App.~\ref{appendix_approx}, which outlines the validity of this approximation). 
The fraction of the mass that actually ends up in the black hole is a very interesting topic, and we leave this to a future study. 

We express the core mass at LS transition through the relation in Eq.~\eqref{eq:v_rho_LS}.
Using Eq.~\eqref{v_core} we express the core mass at the moment of $\gamma=10$ as
\begin{equation}\label{m10}
        M_{c,10}=\sqrt{\frac{6}{\pi}}\left(\frac{1+\xi}{G}\right)^{3/2}\frac{v_{c,10}^3}{\sqrt{\rho_{c,10}}}.
\end{equation}    
The mass at relativistic instability is thus approximately given by 
\begin{equation}\label{bh_mass} 
    M_{c,\rm RI}\simeq M_{c,10}\left(\frac{v_{c,10}}{v_{c,\rm RI}}\right)^{2}\,,
\end{equation}
where $v_{c,\rm RI}\simeq c/3$. 
To demonstrate the relevant mass scales, we use our semi-analytic recipe (in \S\ref{mcore_recipe}) and focus on two example SIDM models, both having $\sigma_0K_p=100\;{\rm cm^2\,g^{-1}}$ at $v=20\; {\rm km\,s^{-1}}$. The models demonstrate the two extremes of velocity dependence; the first has a velocity scale of $w=1 \; {\rm km\,s^{-1}}$ displaying strong velocity dependence, while the second, with $w=10^4 \; {\rm km\,s^{-1}}$, which is practically a constant cross section. We utilize the Python package \textsc{colossus}~\citep{Diemer:2018ApJS..239...35D} for cosmological computations, along with a tight relation between the virial mass of a halo, and its concentration (where the concentration is defined as ratio of the virial radius to the scale radius of a halo) referred to as the concentration-mass relation (see for example \cite{Bullock_2001MNRAS.321..559B,Duffy_2008MNRAS.390L..64D,Klypin_2011ApJ...740..102K} for a detailed account of the relation between concentration and mass of halos). We specifically use the concentration-mass relation from~\cite{Diemer:2019ApJ...871..168D} to compute $M_{c,\rm RI}$, $M_{c,\rm LS}$ and $t_{\rm coll}$ using our recipe in \S\ref{mcore_recipe} and illustrate our findings in Fig.~\ref{fig:mbh_m200_relation}. 

The left panel of Fig.~\ref{fig:mbh_m200_relation} shows the core mass at relativistic instability as a function of $M_{200}$, providing an estimated lower bound on the mass of the black hole formed from core collapse. The middle panel shows the core mass at the LS transition as a function of $M_{200}$, representing the estimated upper bound on the black hole mass. The right panel shows the collapse time (with the collapse time as defined in Eq. 29 of \citetalias{Outmezguine_2022}) of both models as a function of $M_{200}$, with the gray hatched shaded region representing timescales longer than the age of the universe, and thus, signifying halos that do not collapse in time. The width of the bands (shaded purple) in all three panels represents a spread around the median concentration (dark purple line) of 0.3 dex. 
For the left and middle panels of Fig.~\ref{fig:mbh_m200_relation}, the upper (lower) bound of the bands correspond to higher (lower) concentrations, while in the right panel, the upper (lower) bound of the bands represent lower (higher) concentrations, which is indicative of a more rapid collapse for halos with higher concentrations. The concentration dependence of all these parameters (as well as the $M_{200}$ dependence of $M_{c,\rm LS}$ and $M_{c,\rm RI}$) can be found in Appendix~\ref{concentration_dep}. 

The left panel of Fig.~\ref{fig:mbh_m200_relation} shows that higher $M_{c,\rm RI}$ are produced for all halo masses in the case of constant cross sections. 
What is particularly intriguing to note is that, given a high enough concentration, all halo masses with constant cross sections collapse within the age of the universe. Halos with highly velocity-dependent cross sections (namely $w=1\,\rm km s^{-1}$) and masses over $\sim10^9\,M_\odot$ do not collapse within the age of the universe, although a high concentration allows for halos up to $\sim10^{10}\,M_\odot$ to collapse in time. In the case of a constant cross-section, halos of mass $M_{200}\sim10^{10}-10^{11}$ produce a minimum black hole mass ($M_{c,\rm RI}$) of order $1-10^3\,M_\odot$, while highly velocity-dependent ones of the same mass range do not collapse within the age of the universe. This shows that previous works have overestimated the core mass at the relativistic instability by about 2 orders of magnitude.
One should keep in mind, however, that if large cross sections are required on dwarf scales, then models with constant cross sections have been ruled out by cluster constraints~\citep{Kaplinghat:2015aga,Tulin_2018,Sagunski_2021}. Another feature to note is that $M_{c,\rm RI}$ is almost linearly proportional to $M_{200}$ in the highly velocity-dependent case, yielding a nearly straight band in the left panel of Fig.~\ref{fig:mbh_m200_relation}. This dependence on $M_{200}$, along with the relation in the middle panel, can be found in Appendix~\ref{concentration_dep}.

\subsection{Semi-analytic recipe for obtaining the core mass at the relativistic instability}\label{mcore_recipe}

With the tools presented in this paper, one can easily find the core mass of a halo using just the initial halo parameters and the particle physics of the halo. Here, we detail a step-by-step recipe for obtaining $M_c$ at late times, and thus, an estimate for the core mass at the relativistic instability, the minimum mass available for black hole formation.

\begin{enumerate}
    \item Find the maximal core parameters from the NFW halo parameters using the relations from \citetalias{Outmezguine_2022}: 
    \begin{equation*}
        \rho_{c,0}=2.4\rho_s,\,\,\,\,\, v_{c,0} = 0.64V_{\rm max} 
    \end{equation*}
    \item Find the dispersion at LS transition, $v_{c,\rm LS}$. 
    \begin{enumerate}
       \item Use Eq.~\eqref{eq:sigma_hat} to obtain the dimensionless cross section.
        \item Set $\alpha=2.2$ and use Eq.~\eqref{eq:v_rho_LS} to find $v_{c,\rm LS}$. ($\alpha$ typically ranges from $2.19 - 2.22$ for $n_{c,0}$ from $0 - 3.7$, but 2.2 is a sufficient approximation.)
    \end{enumerate}
    \item With the $v_{c,\rm LS}$ obtained in step (ii), use Eq.~\eqref{eq:n_func} to find $n_{c,\rm LS}$. 
    \item Use Eq.~\eqref{eq:v_rho_LS} to find the central density at the LS transition, $\rho_{c,\rm LS}$, and the core mass at LS transition, $M_{c,\rm LS}$.
    \item With the $v_{c,\rm LS},\,\rho_{c,\rm LS},\,n_{c,\rm LS}$ found using the steps above, use our fit in Eq.~\eqref{nls_fits} to obtain the central dispersion and density at the $\gamma=10$ transition,  $v_{c,10},\,\rho_{c,10}$.
    \item Find the core mass at the $\gamma=10$ transition using our equation in Eq.~\eqref{m10}.
    \item Finally, using the second equation in Eq.~\eqref{m10_v10_scaling}, extrapolate to the relativistic instability at $v_c\sim c/3$, and find the minimum mass available for black hole formation using Eq.~\eqref{bh_mass}.

\end{enumerate}
With these seven steps, one can determine the core mass at the relativistic instability, the minimum mass available for black hole formation, simply from knowing the NFW parameters and the assumptions for the particle physics model. See Fig.~\ref{fig:mbh_m200_relation} for the band we obtained using theconcentration-mass relation with \textsc{Colossus} \citep{Diemer:2018ApJS..239...35D} and this recipe.

\section{Conclusions} \label{sec:conclusions}
We have examined the SMFP evolution of spherical, initially NFW, isolated halos for velocity-dependent SIDM with varied velocity dependence strength undergoing elastic scattering only. We show that in the SMFP regime, the core becomes distinct from the outer LMFP halo.
For the first time, we discover a universality in the SMFP regime and the appropriate timescale in the SMFP evolution when the halo is deep in the core collapse regime. We find new scales that allow one to scale out the particle physics, specifically at the $\gamma=10$ transition which leads into the constant thermal energy phase in Stage 3 of the halo evolution (see Fig.~\ref{fig:log_slope_tscale}). It remains unclear why this universality exists, but we find it in all our runs for velocity-dependent SIDM halos. 
Although the $\gamma=10$  transition point cannot be found analytically, we have determined a relation between the LS and $\gamma=10$ transitions given in Eqs.~\eqref{nls_fits} that allows to simply compute parameters in Stage 3 given initial halo parameters and particle physics. 

With these findings, we devise a recipe to estimate the core mass deep in the core collapse regime using the relation in Eq.~\eqref{eq:v_rho_LS}, shown in Fig.~\ref{fig:mc_rhoc_vc_plots}, and our recipe in \S\ref{mcore_recipe}. We can use this recipe to compute the core mass at  relativistic instability, which we expect to be the minimum mass available for black hole formation, given the initial halo parameters and particle physics. We find that previous works based on the gravothermal solutions have overestimated the core mass at relativistic instability by about 2 orders of magnitude. We find that
the mass of the core reaches $\sim10^3\,M_\odot$ at the relativistic instability only for $M_{200}\sim10^{11}\,M_\odot$ and constant cross sections, while high velocity dependence prevents collapse from occurring in more massive halos. However, we note that other particle physics likely becomes important in this regime and may cause black hole formation to occur before the halo reaches $v_c\sim c/3$. 
Note that dissipation \citep{Essig:2018pzq,Xiao:2021JCAP...07..039X,Feng_2022JCAP...05..036F}, halo truncation \citep{Nishikawa_2020}, and the presence of baryons \citep{Feng_2021} have been shown to cause halos to collapse faster, which could lead to larger black hole masses.

We compute the mass within the region of the halo that is in the SMFP regime as a function of time, and show that it asymptotes to a constant value. We show that this constant mass value can be estimated using the mass within the core when the core enters the SMFP regime to within an $\mathcal{O}(1)$ scatter (see App.~\ref{appendix_approx}. The core mass at the LS transition can be easily estimated given the analytic approximations derived in~\citetalias{Outmezguine_2022}. We conjecture that this quantity is roughly the maximum mass available for black hole growth via other physical processes, such as accretion.

The key takeaways of our work are:
\begin{itemize}[leftmargin=*, labelindent=1pt, itemsep=.2cm, parsep=0pt]
    \item \textit{We have discovered a universal solution for halos in Stage 3 of the evolution (see Fig.~\ref{fig:log_slope_tscale}).}
    \item \textit{We have devised a semi-analytic method to determine the $\gamma=10$ transition parameters by relating them to the LS parameters defined in \citetalias{Outmezguine_2022}.}
    \item \textit{We have outlined a step-by-step recipe to compute the properties of the core deep in the core collapse regime, given the halo parameters and cross section for models with elastic scattering. As an application, we determine the core mass at the relativistic instability, which serves as the minimum mass available for black hole formation.}
\end{itemize}
Coupled with our companion paper~\citetalias{Outmezguine_2022}, we now have appropriate relations to describe the entire gravothermal evolution of isolated halos, from the LMFP to SMFP regimes. This provides a simple way to compute the minimum core mass available for black hole growth after core collapse, given the initial halo parameters and assuming no physics other than elastic collisions. It is an interesting question to ask how these semi-analytic predictions would change for tidally truncated halos and whether the universality based on the connection to the LMFP regime is retained to some extent. It would also be interesting to extend our analysis to cases where baryons dominate the central potential well and where dissipative interactions are important.

\section*{Acknowledgements}
MK and LS thank the Pollica Physics Center, where part of this research was carried out, for its warm hospitality. The Pollica Physics Center is supported by the Regione Campania, Università degli Studi di Salerno, Università degli Studi di Napoli ``Federico I'', the Physics Department ``Ettore Pancini'' and ``E.R. Caianiello'', and the Istituto Nazionale di Fisica Nucleare.

KB acknowledges support from the National Science Foundation (NSF) under Grant No.~PHY-2112884. MK acknowledges support from the NSF under Grant no.~PHY-2210283. The work of NJO was supported in part by the Zuckerman STEM Leadership Program and by the NSF under the grant No.~PHY-1915314. The work of SGN was supported by the Departmental Dissertation Fellowship of the Department of Physics and Astronomy at the University of California, Irvine.

\section*{Data Availability}

Our gravothermal code is available at \url{https://github.com/kboddy/GravothermalSIDM}. For a direct example of how to apply our recipe in \S\ref{mcore_recipe}, a sample script is available at \url{https://github.com/Nadav-out/Gravothermal-Instability-mass}.
The data for all the runs in this paper are available from the corresponding author upon request.



\bibliographystyle{mnras}
\bibliography{SMBH} 



\appendix

\section{Semi-analytic Description of the Evolution of the Core} \label{appendix_derivations}
In this appendix, we derive some semi-analytical results, describing the central halo evolution in the deep SMFP regime. We start our derivation by focusing the short-distance behavior of the gravothermal solution. To do so, we Taylor expand near $r=0$ the halo density and velocity dispersion profiles as\footnote{while the ${\cal O}(x^4)$ term in the expansion of $\rho$ is of similar magnitude to that of $v^2$, we don't quote it here as it will not appear in any future steps.}
\begin{equation}\label{Jeans_eq}
    \begin{split}
	\rho &= \rho_c\left(1-x^2+ {\cal O}(x^4)\right), \\ v^2 &= v_c^2\left(1-\xi x^2 +\xi\beta x^4 + {\cal O}(x^6) \right),
    \end{split}
\end{equation}
where $x=r/r_c$, and $r_c$ is defined through 
\begin{equation}
    r_c^{-2}=-\lim_{r\to0}\left(\frac{\partial \log\rho}{\partial r^2}\right)_t.
\end{equation}
Note that $\xi$ can be defined thorough
\begin{equation}
    \xi=\lim_{r\to 0}\left(\frac{\partial \log v^2}{\partial\log\rho}\right)_t.
\end{equation}
Mass conservation gives that 
\begin{equation}
    M=M_c\left(x^3+{\cal O}(x^5)\right)\;\;,\;\;M_c=\frac{4\pi}{3}\rho_cr_c^3.
\end{equation}
Using this in Jeans' equation we arrive at the relation
\begin{equation}\label{eq:central_quants}
    v_c^2=\frac{G M_c}{2r_c(1+\xi)}.
\end{equation}
To make use of the remaining gravothermal equations, we recall Eqs.~\eqref{eq:kappa_smfp} and~\eqref{eq:n_func}, which allows us to write
\begin{equation}
    \kappa_{\rm SMFP}(v)=\kappa_{\rm SMFP}(v_c)\left(1-\frac{n+1}{2}\xi x^2+{\cal O}(x^4)\right).
\end{equation}
With this, the derivative of luminosity at short distances is given by
\begin{equation}\label{eq:dL_dr_1}
    \frac{\partial L}{\partial r}\simeq8\pi v^2_c m_{\rm dm}\kappa_{\rm SMFP}(v_c)\xi x^2\left[3-\frac{5}{2}x^2\left(4\beta+(n+1)\xi\right)\right].
\end{equation}
The above equation should be equated to the entropy conservation law of the gravothermal equations, Eq.~\eqref{eq:gravothermal}. To do so, we first note that, using properties of partial derivative, we find that at small $x$ the following holds true
\begin{equation}
    \left(\frac{\partial f}{\partial t}\right)_M \simeq\left(\frac{\partial f}{\partial t}\right)_x-\frac{x}{3} \frac{d \log M_c}{d t}\left(\frac{\partial f}{\partial x}\right)_t.
\end{equation}
Therefore
\begin{equation}
    \left(\frac{\partial}{\partial t}\right)_M\log\frac{v^3}{\rho}\simeq D_1(t)-\frac{x^2}{3}(2-3\xi)D_2(t)
\end{equation}
with $D_1=d\log( v_c^3/\rho_c)/dt$ and $D_2=d\log M_c/dt$. With this, entropy conservation gives
\begin{equation}\label{eq:dL_dr_2}
    \frac{\partial L}{\partial r}\simeq-4\pi r_c^2\rho_cv_c^2x^2\left\{D_1-\frac{x^2}{3}\left[3(1+\xi)D_1+(2-3\xi)D_2\right]\right\}
\end{equation}
Equating to leading order the two expressions we have for $dL/dr$ in Eqs.~\eqref{eq:dL_dr_1} and~\eqref{eq:dL_dr_2} we get
\begin{equation}\label{eq:d_log_Sc_dt}
    \frac{d\log v_c^3/\rho_c}{dt}=-\frac{3}{t_{\rm SMFP}}\frac{\xi}{1+\xi}\;\;,\;\;t_{\rm SMFP}=\frac{\sigma_0K_{\rm eff}^{(2)}(v_c)v_c}{2\pi b G m_{\rm dm}}.
\end{equation}
Moving to the next order in $dL/dr$ we find
\begin{equation}\label{eq:d_log_Mc_dt}
   \left(\frac{2}{3}-\xi\right) t_{\rm SMFP}\frac{d\log M_c}{dt}=\left(3-10\beta+\frac{1-5n}{2}\xi\right)\frac{\xi}{1+\xi}
\end{equation}

We recall that in the deep SMFP regime, we find that $\rho_c\propto v_c^\gamma$, with $\gamma\simeq 10$. Through Eq.~\eqref{eq:central_quants}, this also implies $M_c\propto v_c^{3-\gamma/2}$. With this in mind, the ratio of Eqs.~\eqref{eq:d_log_Sc_dt} and~\eqref{eq:d_log_Mc_dt} allows to relate $\gamma$ to $\xi, \beta$ and $n$
\begin{equation}
   \xi=\frac{6+(20\beta-8)(\gamma-3)}{9-(2+5n)(\gamma-3)}\simeq\frac{10-28\beta}{1+7n}
\end{equation}

\section{The dependence of the SMFP parameters on the concentration}\label{concentration_dep}
In this section of the appendix, we outline the dependence of the SMFP parameters, such as the parameters at the LS transition, the $\gamma=10$ transition, and the black hole mass, on the concentration, at fixed $M_{200}$. 

To show how the parameters depend on the concentration, it is useful to outline these relations first:

\begin{equation}
    \begin{split}
        \frac{r_{200}}{r_s} &= c_{200}\;\;, \\
        \frac{\rho_s}{\rho_{\mathrm{crit}}} &= \frac{200}{3}\frac{c_{200}^3}{\log (1+c_{200})-\frac{c_{200}}{1+c_{200}}}\;\;.
    \end{split}
\end{equation}
(see for example \cite{Navarro_1996}). Given that our parameters depend on some combination of $\rho_s,V_{\mathrm{max}}$, we can go ahead and find the dependence on the concentration with fixed $M_{200}$. 

It is right away useful to note that the dependence of $\rho_s$ on $c_{200}$ approaches $c_{200}^3$, but for the range of $c_{200}$ we used in our spread, the power ranges between $p = 2.4-2.6$. We also note that $V_{\rm max}\propto\sqrt{\rho_s r_s^2}$ which allows us to find the dependence of $V_{\rm max}$ on $c_{200}$.

We now note that the LS transition parameters are related to $\rho_s,V_{\rm max}$ as follows:

\begin{equation}
    \begin{split}
        v_{c,\rm LS}&=v_{c,0}\hat{\sigma}_{c,0}^{-1/\delta}\propto V_{\rm max}\hat{\sigma}_{c,0}^{-1/\delta}\;\;,\\
        \rho_{c,\rm LS}&=\rho_{c,0}\frac{v_{c,\rm LS}}{v_{c,0}}^\frac{2\alpha}{\alpha-2}\propto \rho_{s}\left(\hat{\sigma}_{c,0}^{-1/\delta}\right)^\frac{2\alpha}{\alpha-2}\;\;,\\
        \hat{\sigma}&\propto K_n\left(\frac{v_{c,0}}{w}\right)v_{c,0}\sqrt{\rho_{c,0}}\propto v_{c,0}^{n+1}\sqrt{\rho_{c,0}}\;\;,\\
        v_{c,10} &\propto v_{c,\rm LS}\;,\;
        \rho_{c,10}\propto\rho_{c,\rm LS}\;,\\
        M_{c,10}&\propto\frac{v_{c,10}^3}{\sqrt{\rho_{c,10}}}\;,\;
        M_{c,\rm RI}\propto M_{c,10}v_{c,10}^2\;\;.
    \end{split}
\end{equation}
We now see how nicely all the SMFP parameters are related to $V_{\rm max},\rho_s$, and we know how these parameters are related to $c_{200}$. This gives us the dependence on the concentration for these parameters, as compared to these parameters at the median concentration, as follows:
\small
\begin{equation}\label{c_dep}
    \begin{split}
        \log\left(\frac{v_{c,10}}{v_{c,\rm 10(med)}}\right)&= d\left[q\left(1-\left\{\frac{1-n}{\delta}\right\}\right)-\left\{\frac{q+1}{\delta}\right\}\right]\,\\
        \log\left(\frac{\rho_{c,10}}{\rho_{c,\rm 10(med)}}\right)&= 2 d\left[-q\eta\left\{\frac{1-n}{\delta}\right\}+(\delta-\eta)\left\{\frac{q+1}{\delta}\right\}\right]\,\\
        \log\left(\frac{M_{c,10}}{M_{c,\rm 10(med)}}\right)&= d\left[q\left(3+(\eta-3)\left\{\frac{1-n}{\delta}\right\}\right)-(3-\eta+\delta)\left\{\frac{q+1}{\delta}\right\}\right]\,\\
        \log\left(\frac{M_{c,\rm RI}}{M_{c,\rm RI(med)}}\right)&= d\left[q\left(5+(\eta-5)\left\{\frac{1-n}{\delta}\right\}\right)-(5-\eta+\delta)\left\{\frac{q+1}{\delta}\right\}\right]\,\\
        \log\left(\frac{M_{c,\rm LS}}{M_{c,\rm LS(med)}}\right)&=\\
        d&\left[q\left(3-(3-2\eta)\left\{\frac{1-n}{\delta}\right\}\right)-(3-2\eta+\delta)\left\{\frac{q+1}{\delta}\right\}\right]
    \end{split}
\end{equation}
\normalsize
where here, $d=\pm0.3$ for the spread in the concentration, $n=n_{c,0}$, $\delta$ is given by Eq.~\eqref{eq:v_rho_LS}, we have set $\eta=\frac{\alpha}{\alpha-2}$ with $\alpha=2.2$ being a good approximation and thus $\eta=11$, $q=\frac{p}{2}-1$ and $p=2.4-2.6$ where the lower bound of $c_{200}$ is $p=2.4$, and the upper bound is $p=2.6$ (but $p=2.5$ is sufficient). This gives us the dependence on the concentration of the SMFP parameters {\it at fixed $M_{200}$}. One may also find the dependence on $M_{200}$ by fixing $c_{200}$. For example, since $V_{\rm max}\propto M_{200}^{1/3}c_{200}^{p/2-1}$, we find that
\begin{equation}\label{m200_dep}
    \begin{split}
        \log\left(\frac{M_{c,\rm RI}}{M_{c,\rm RI(med)}}\right)= \frac{1}{3}\left(5+(\eta-5)\left\{\frac{1-n}{\delta}\right\}\right)\,,\\
        \log\left(\frac{M_{c,\rm LS}}{M_{c,\rm LS(med)}}\right)= \frac{1}{3}\left(3+(3-2\eta)\left\{\frac{n-1}{\delta}\right\}\right)\,,
    \end{split}
\end{equation}
which easily explains the behavior we see in the left panel of Fig.~\ref{fig:mbh_m200_relation} of the ratio $\frac{M_{c,\rm RI}}{M_{200}}$ where the $w=1\,\rm km/s$ model (which has high $n$) appears to be relatively constant, and indeed this is because its dependence on $M_{200}$ is such that $M_{c,\rm RI}\propto M_{200}$ as $n\rightarrow 3.7$, and closer to quadratic as $n\rightarrow 0$.

\section{Resolution}
\label{resolution}
In this section, we discuss the numerical resolution issues encountered when probing the SMFP regime. In this regime, the central density increases drastically, while the core size shrinks significantly as well. This results in the core size becoming too small to be resolved late in the evolution. To address this while maintaining reasonable run times, we probe smaller radii than we did in \citetalias{Outmezguine_2022} with some trade-off to maintain reasonable run times of the code, and increase the number of shells so as not to sacrifice the resolution of each profile. We also develop a method for finding where the central density is no longer resolved by finding the snapshot at which the core has become too small to resolve, and place a cutoff to exclude snapshots with profiles where the core can no longer be resolved.

\subsection{Probing smaller radii and increasing shells}
As halos evolve into the SMFP, the central density experiences a drastic increase while at the same time the size of the core shrinks significantly. This eventually will lead to the core shrinking to radii smaller than those we are probing. The solution is to examine smaller radii while maintaining reasonable run times. To accommodate this while maintaining reasonable run times for the code, we probed smaller radii in our runs ($\Tilde{r}_\tx{min}=0.005$ compared to $\Tilde{r}_\tx{min}=0.01$ in \citetalias{Outmezguine_2022}), and increased the resolution from 400 to 450 shells to ensure our shell resolution was not compromised. 

Even with this increased resolution and probing smaller radii, the SMFP evolution will drive the cores of the halos to shrink beyond the resolution we set in our runs. The reason this becomes problematic is that we are not actually taking the density at the center of the halo ($r=0)$, and as such the central density $\rho_c$ we obtain numerically is an approximation that depends on how small our innermost radial shell is. If the core of the halo is resolved, then the central region of the profile will be approximately flat, and thus our $\rho_c$ can be trusted to be a good approximation. But as the core will shrink to a point where it can no longer be resolved, a test is required to ensure we only probe snapshots with resolved cores, and remove those that are not from the analysis.

\subsection{Core resolution test \label{sec:Core_resolution_test}}
To address the issue of core resolution in our snapshots, we develop a test to determine whether or not the core is resolved, and thus, whether the approximate central density can be trusted. To do so we estimate the log-slope of the density profile at fixed mass $s[n] = (\ln(\rho[n+1]/\rho[n-1])/\ln(r[n+1]/r[n-1])$ from the $n+1$ and $n-1$ points. We place a stringent limit on the slope that determines when the snapshots are no longer resolved, such that if the slope at the nth point is steeper than -0.5, then we cannot assume the core is resolved any longer and remove this and later snapshots from any analysis where the core density is required. We take $n$ to be the fourth point in the density profile of a halo at each snapshot in time in order to avoid numerical artifacts in the initial points in the profile.

One of the quantities of interest for our analysis is the rate of change of the central density with the central velocity dispersion, i.e., $\gamma=d\ln(\rho_c)/d\ln(v_c)$. We estimate this as $\gest[n]$ using the density $\rho[n]$ and $v[n]$ at different snapshots in time. Note that the central velocity dispersion tends to a constant much faster than the density profile, and so $v[n]$ is close to $v_c$ to a good approximation. 
Since our derivatives are computed at fixed mass, we can ascertain how good an approximation $\gest[n]$ is. Within the core, we can write $\rho[n] = \rho_c(1-(r[n]/r_c)^2)$ for $r[n] \ll r_c$, and this ansatz defines $r_c$. In this region, we write (approximately) the enclosed mass within $r[n]$ as $M[n] = (4\pi/3)\rho_c r[n]^3$. Within this set of approximations, we can show the following:   
\begin{equation}\label{eq:g_approx}
        \gamma=\frac{\gest[n]-s[n]}{1-s[n]/6},
\end{equation}
where $\gest[n]$ is shown in Fig.~\ref{fig:log_slope_tscale}. By limiting to $s[n] > -0.5$, we can see that the error is only about $3.3\%$ or less when $\gest[n] \approx 10$.

\subsection{Stage 3 of the evolution}
Stage 3, the region beyond the $\gamma=10$ transition, cannot directly be obtained analytically. Because of this, we approximate the transition by selecting the minimum of the curve in Fig.~\ref{fig:log_slope_tscale} (left panel; diamonds correspond to the local minimum, which we call the $\gamma=10$ transition). 
This is the point beyond which the slope begins to increase, during which the halos are in Stage 3. It is visible that the transition point is not where the curves reach the $\gamma=10$ solution just yet; the models with weaker velocity dependence (or smaller $n$) take slightly longer to asymptote to the $\gamma=10$ solution (i.e. Stage 3), as can be seen in the left panel of Fig.~\ref{fig:log_slope_tscale}. For smaller values of $n<1$, 
we cannot numerically resolve much of Stage 3; in fact, these halos fall to $\gamma=3$. However, for these models, we also observed that once in this $\gamma=3$ region, the density of the shells just outside the core changes drastically from shell to shell; because the gravothermal equations have been discretized assuming changes from shell to shell are small, the results in this region cannot be trusted. We have also explicitly checked that in these cases where the halos fall to this $\gamma=3$ region, energy is not conserved. This leads us to the conclusion that the $\gamma=3$ region is numerical, and not an actual solution. It seems that the models with $n<1$ suffer this issue because the core mass is the largest for these models. We hypothesize that numerical effects are larger for halos with larger cores (thus, those with models that have $n<1$). A detailed analysis of this is beyond the scope of this paper, and we thus simply acknowledge that there are numerics involved here, but do not investigate further. We thus caution the reader that, for $n<1$, we rely on assumptions and the evolution observed for the  $n\geq1$ cases, which indicate that the halos evolve to asymptote to the $\gamma=10$ region in Stage 3.

Despite using the local minimum as the $\gamma=10$ transition point (as opposed to, say, a point on the asymptote), we find that selecting $t_{\rm SMFP}(t_{10})$ scales the collapse times of the halos astonishingly well, as can be seen in the right panel of Fig.~\ref{fig:rhoc_t_plots}. Thus, this approximation works well enough to make predictions.

\section{Accuracy of Analytic Predictions and Approximations}\label{appendix_approx}

In this section, we discuss briefly our choice of using the analytic prediction for $v_{c,\rm LS},\rho_{c,\rm LS}$ rather than our numerical results, we justify the truncation of the expansion of Eq.~\eqref{k_secondorder}, we explain our definition of the mass at the transition from the LMFP to the SMFP regime, and we outline the use of a fitted function as an approximation in place of Eq.~\eqref{eq:K_p}. 

\begin{figure*}
  \centering
  \includegraphics[width=0.475\textwidth]{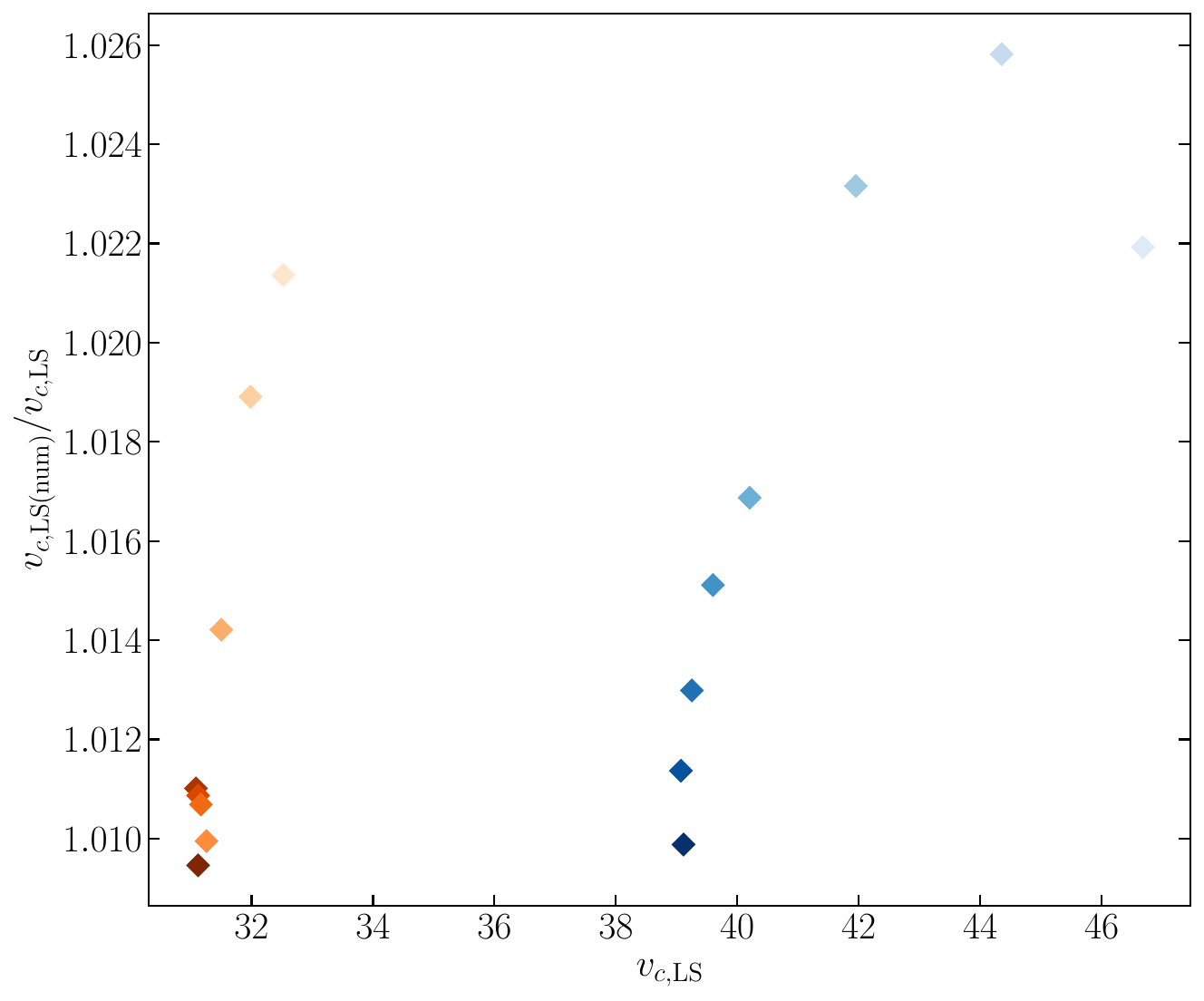}\;\;\;\includegraphics[width=0.472\textwidth]{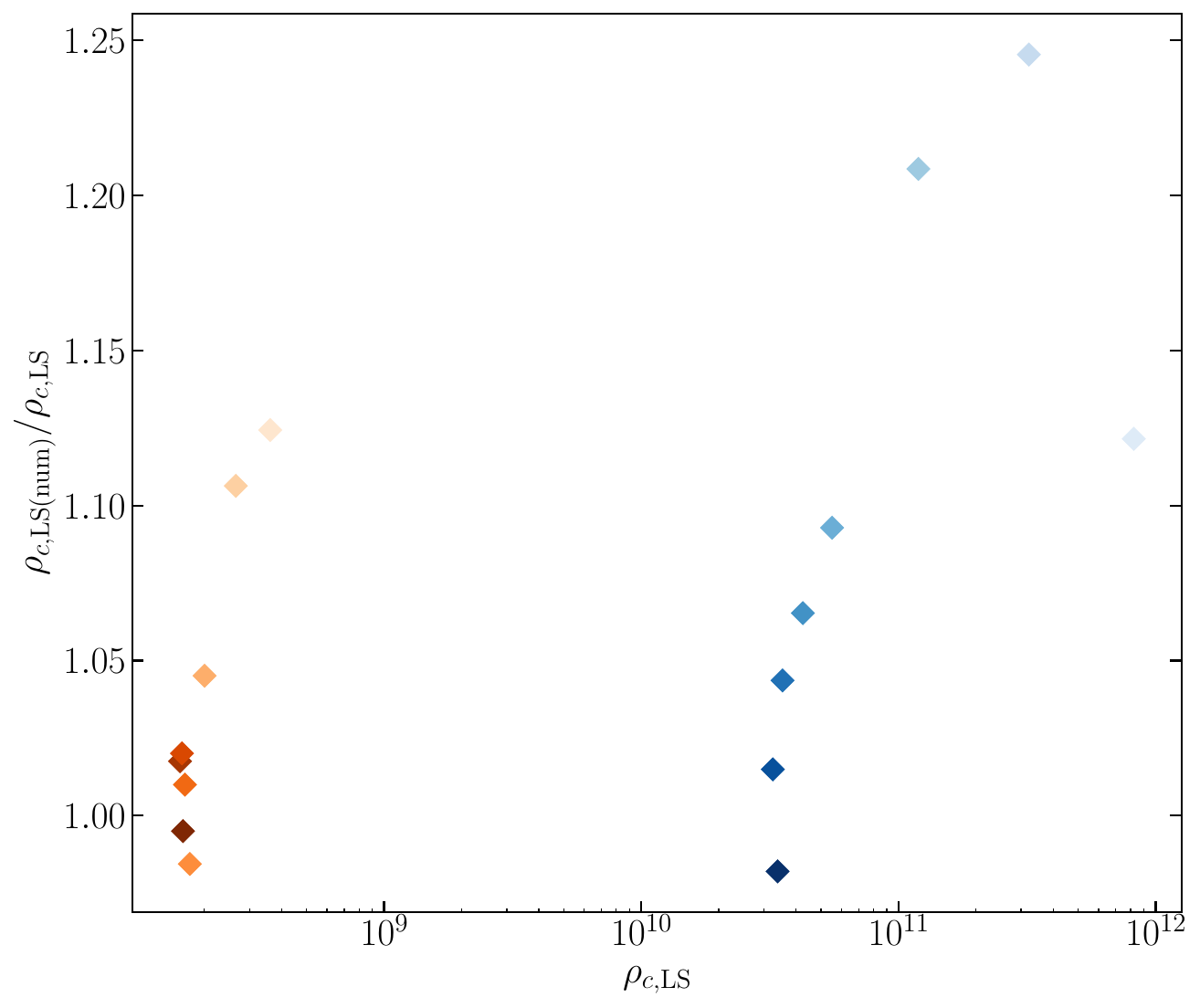} 
  \caption{ {\it (Left)} 
Comparison between the numerically obtained and analytically predicted central velocity at LS transition, plotted as a function of the analytic central velocity at the LS transition. The difference between them is very small, within $\sim2.6\%$. 
 {\it (Right)} 
Comparison between the numerically obtained and analytically predicted central density at LS transition, plotted as a function of the LS transition central density. Here the analytic and numerical values differ by at most $\sim25\%$. It is unsurprising that the central density is a bit less accurate as it is the parameter that changes drastically.} \label{fig:v_rho_ls_pred_num}
\end{figure*}

A core aim of this paper is to provide a method to predict all parameters of a halo at both LS and $\gamma=10$ transitions and black hole masses given only the initial halo parameters and the particle physics model used. To achieve this, it makes most sense to use the prediction for $v_{c,\rm LS}$ in Eq.~\eqref{eq:v_rho_LS} to show that the entirety of the halo evolution can be predicted without the need to solve the gravothermal equations. Comparing the numerically determined and theoretically predicted values, we find the $v_{c,\rm LS}$ prediction is within $\sim2.6\%$ of the numerically determined value, while $\rho_{c,\rm LS}$ is within $\sim25\%$, as seen in Fig.~\ref{fig:v_rho_ls_pred_num}. The wider scatter in the central density can be attributed to the incredibly large changes it undergoes through the evolution. Given that the predictions agree with the numerical values to within 25\%, we conclude that the analytic parameters are sufficiently close to the numerical values and use them throughout this work. In the case of the parameters at the $\gamma=10$ transition, we use the numerically obtained values, followed by fitting a curve to provide a semi-analytic method of obtaining the parameters.

\begin{figure*}
\centering
\includegraphics[width=0.6\textwidth]{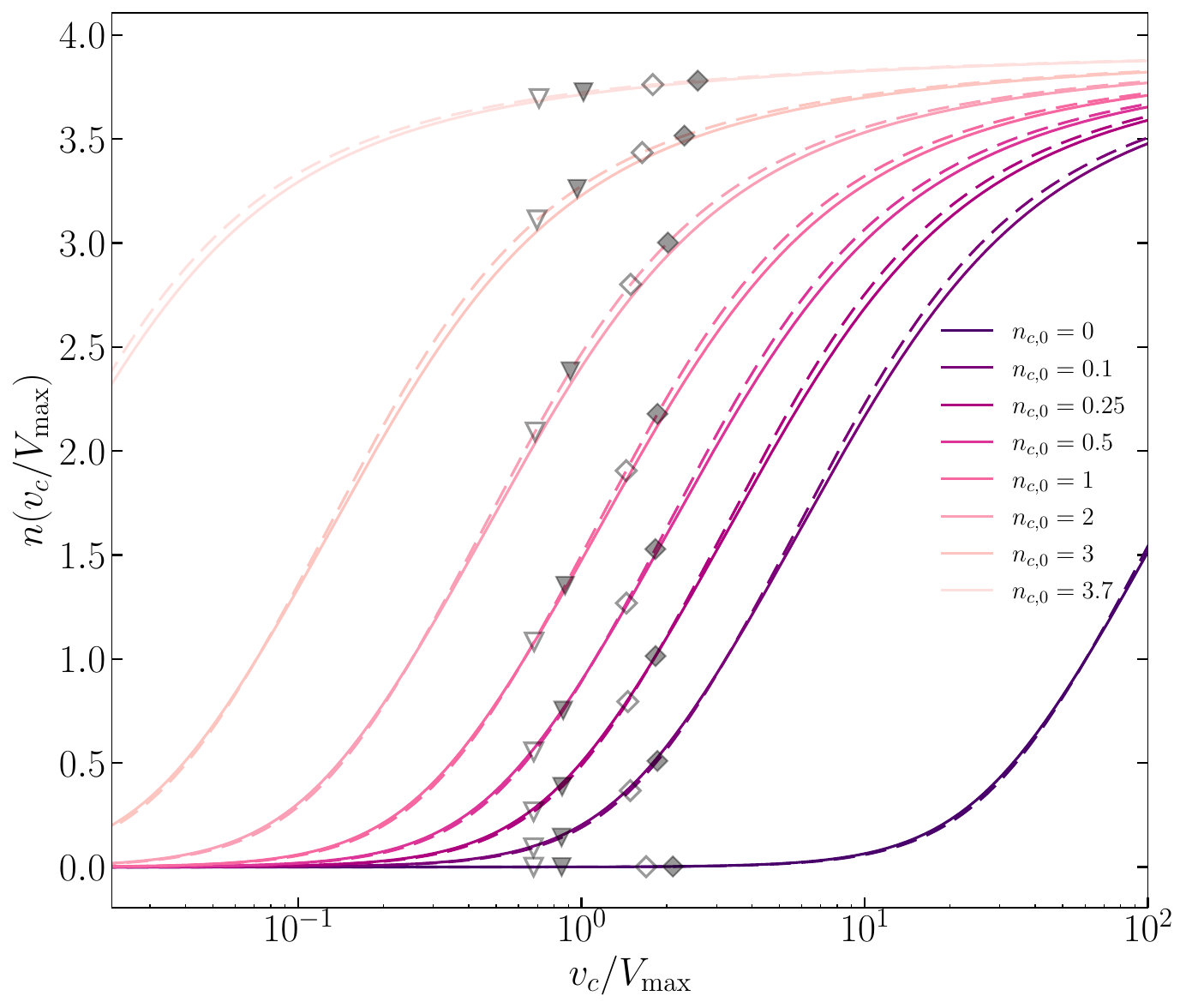}
\caption{ 
Plot of $n$ as a function of $v_c$, plotted by holding the corresponding $w$ constant for each curve. The triangles show the computed $n_c$ and $v_c$ for the LS transition, and the diamonds are for the computed $\gamma=10$ transition. In both cases, the solid markers are for runs 1-8 (lower $\hat{\sigma}$, and open markers for runs 9-16 (higher $\hat{\sigma}$), as shown in Table \ref{table:run_params}. The solid lines are plotted using $K_5$, while the dashed lines use $K_{\rm eff}^{(2)}$.
}
\label{fig:n_vs_vc}
\end{figure*}

Next, we justify the analytic approximation of our second order expansion in Eq.~\eqref{k_secondorder} and our choice for truncating the expansion of $K_{\rm eff}^{(2)}$ at the second order, where we show the difference between $K_5$ and $K_{\rm eff}^{(2)}$. One will note that the curves in Fig.~\ref{fig:n_vs_vc} vary only slightly, 
and the largest variation occurs at high $v_c$ when the halos are in the SMFP. As we said in \S\ref{particlephysics_conductivity}, the variation between $K_5$ and $K_{\rm eff}^{(2)}$ at second order is up to 20\%, while the difference between the second order and the next expansion is only a sub percent correction. As seen in Fig.~\ref{fig:n_vs_vc}, the next order correction would not change the curves very much at all, thus truncating the expansion at second order is valid and sufficient.

The mass within the LS transition radius (that is, the radius at which Eq.~\eqref{kappa_ratio} is satisfied, i.e., mass of a halo that is within the SMFP regime) is an important quantity. This evolves with time and eventually plateaus to a constant value. We conjecture that this would serve as an upper limit on the mass available for black hole growth. This mass cannot be determined analytically. But we find that the core mass at the time the core transitions into the SMFP, $M_{c,\rm LS}$, offers a good approximation for this mass. This is desirable, as we can analytically determine the halo core properties at the LS transition using the relations derived in \citetalias{Outmezguine_2022}. 

\begin{figure*}
  \centering
  \includegraphics[width=0.476\textwidth]{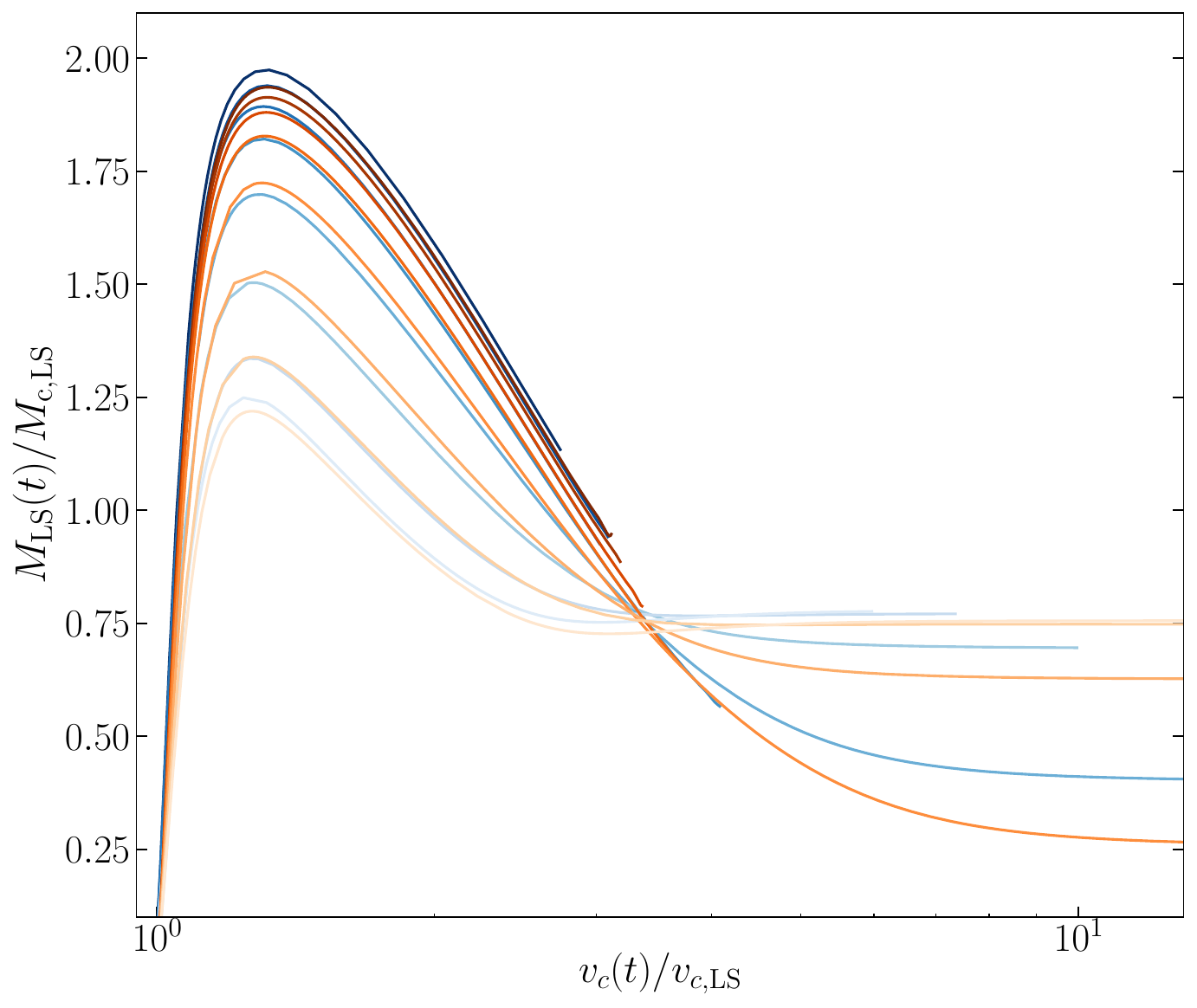}\;\;\;\includegraphics[width=0.47\textwidth]{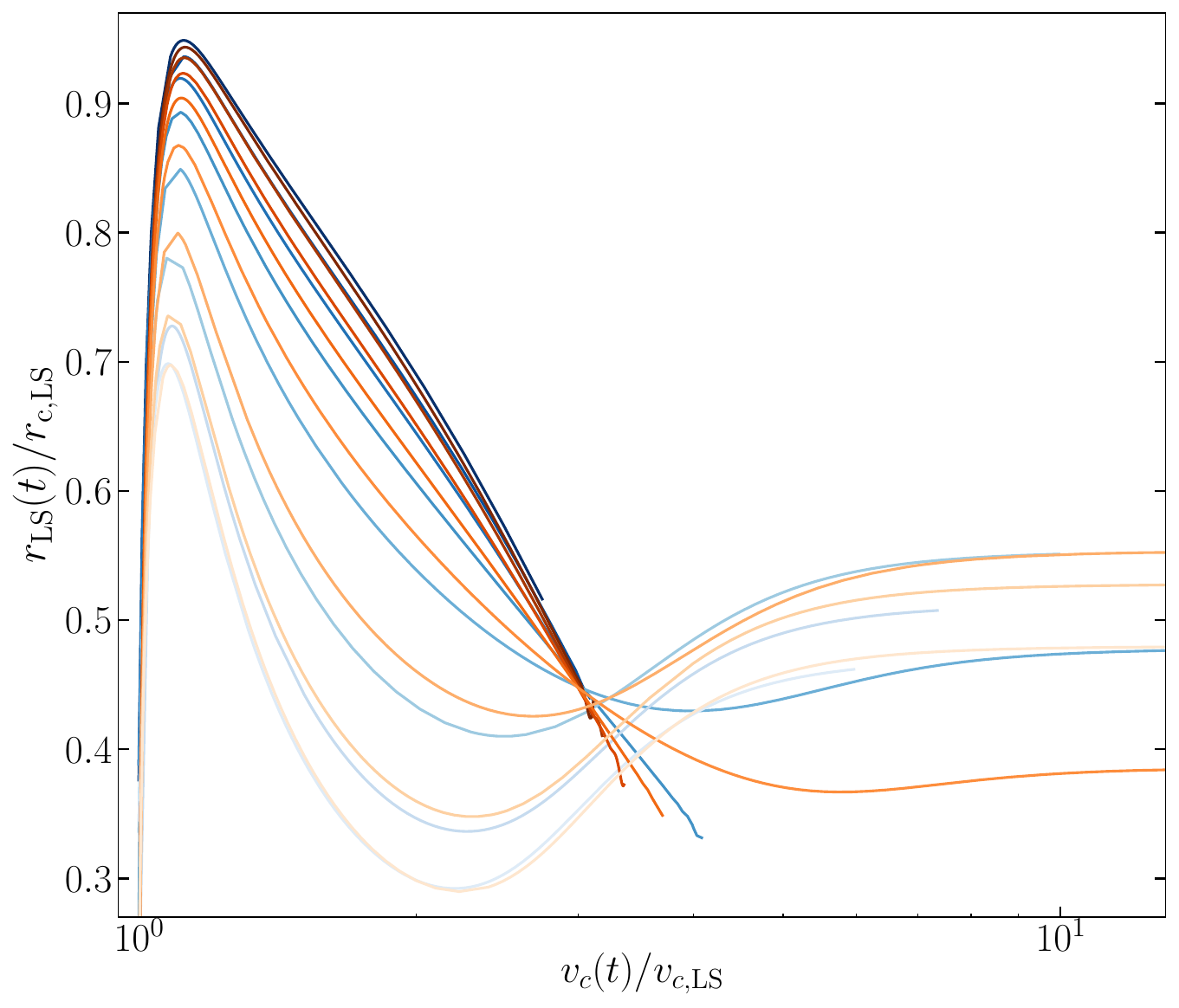}
  \caption{ {\it (Left)} 
The LS transition mass normalized by the core mass at LS transition as a function of central velocity.
 {\it (Right)} 
The LS transition radius normalized by the core radius at LS transition as a function of central velocity.}
\label{fig:m_r_transition_plots}
\end{figure*}

\begin{figure*}
\centering
\includegraphics[width=0.8 \textwidth]{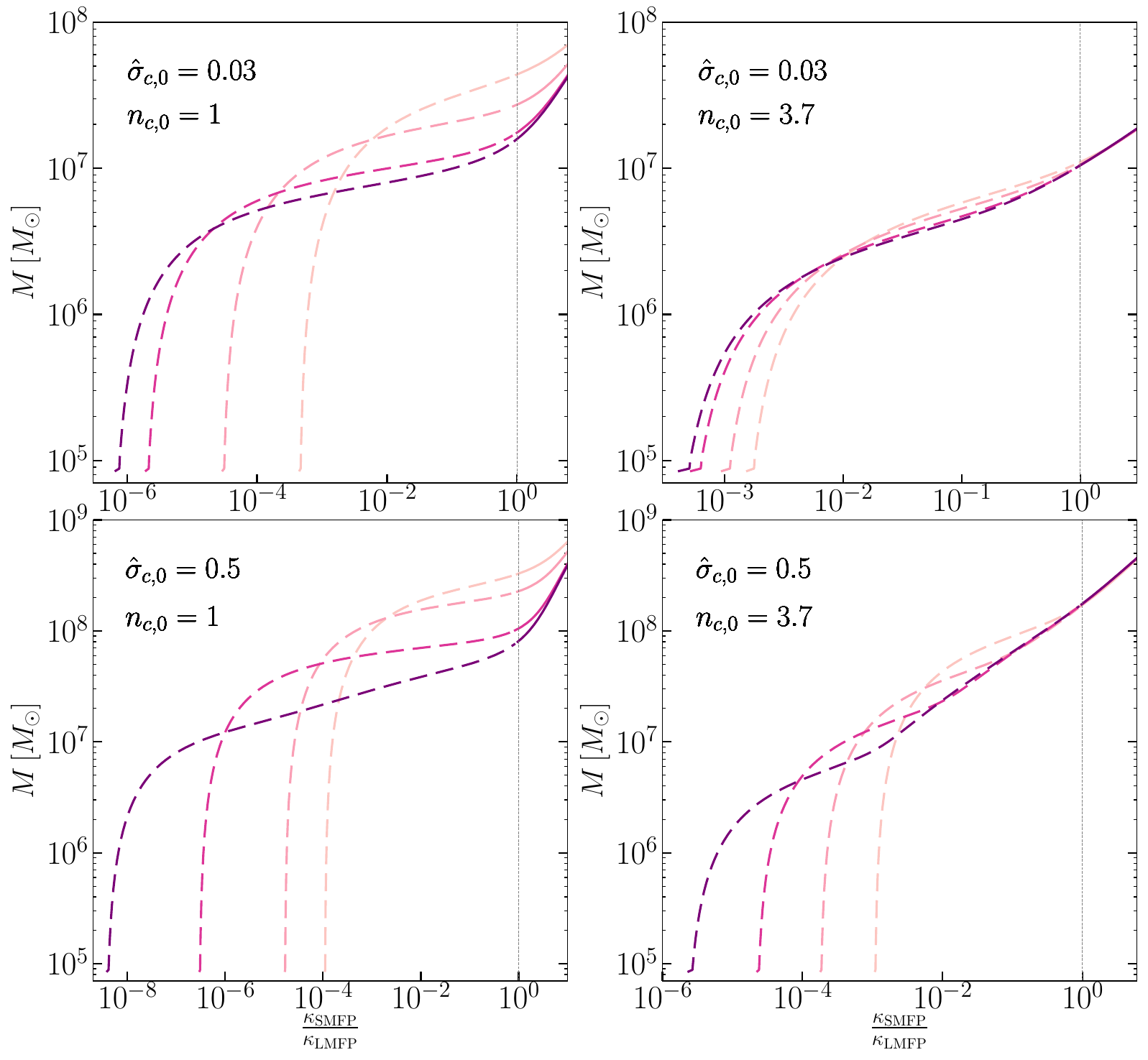}\;\;\;
\caption{ Mass at LS transition as a function of $\frac{\kappa_{\rm SMFP}}{\kappa_{\rm LMFP}}$ for 4 cases, corresponding to Runs 5, 8, 13, 16. The curves each correspond to profiles that are in Stage 3 of the evolution.
}
\label{fig:m_vs_kappa}
\end{figure*}

In Fig.~\ref{fig:m_r_transition_plots}, we 
plot both the ratio of the masses, $M_{\rm LS}/M_{c,\rm LS}$,
and of the radius, 
$r_{\rm LS}/r_{c,\rm LS}$, as a function of the central velocity normalized by the central velocity at the LS transition. We also show the evolution of the mass profiles that are in the SMFP regime as a function of the kappa ratio $\frac{\kappa_{\rm SMFP}}{\kappa_{\rm LMFP}}$, and that are in Stage 3, for four cases that show a range in $n$ and $\hat{\sigma}$ in Fig. \ref{fig:m_vs_kappa}. One can see that for higher $n$, generally the LS mass does not evolve as much as for lower $n$. In Fig.~\ref{fig:m_r_transition_plots} we see that the ratio of the LS mass to the core mass at LS also tends to a constant in the later evolution in Stage 3 (the same is true for the radius). The ratio only differs by $\mathcal{O}(1)$, making the core mass at LS transition valid and more useful to use in place of the mass of a halo at the LS transition, as it can be determine analytically.

\begin{figure*}
\centering
\includegraphics[width=0.6\textwidth]{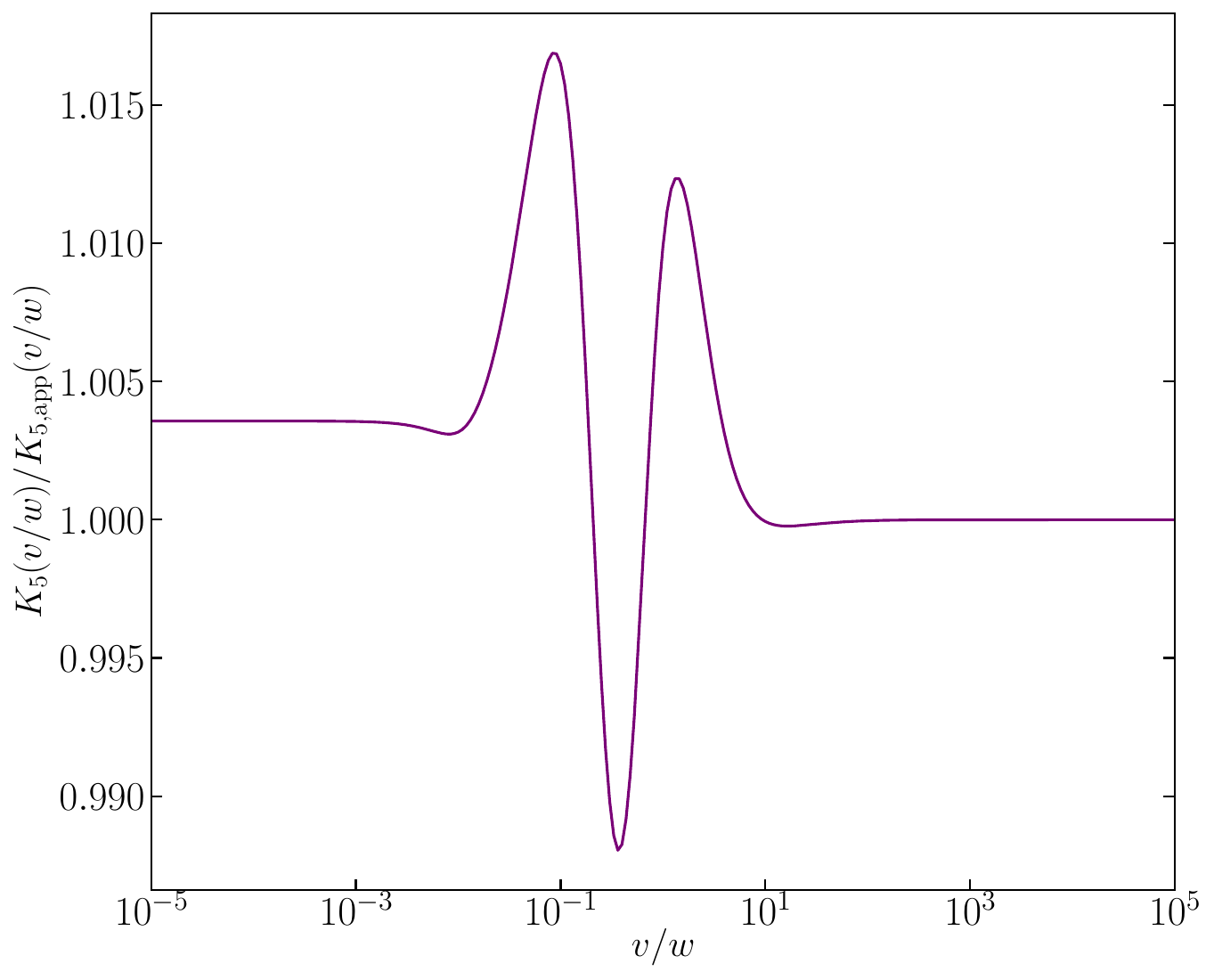}
\caption{ 
The ratio of the exact $K_5$ function as in Eq.~\eqref{eq:K_p} to the $K_5$ approximation given in Eq.~\eqref{eq:K_p_approx} with respect to $v/w$. The functions differ by less than 2\%.
}
\label{fig:Kfunc_comp}
\end{figure*}

Finally, we mention the difference in the way we use Eq.~\eqref{eq:K_p} in the gravothermal code and in the analysis. In the gravothermal code, we used a fit approximation for the purpose of maintaining numerical stability. The approximation is given by

\begin{equation}\label{eq:K_p_approx}
    K_p\approx\left(1+\frac{sp_0p_2}{1.5\log[1+(sp_1)^{p_2}]}\right)^{-1/p_3}\,,
\end{equation}
where we have set $s=x^4+\varepsilon$ with $x=v/w$ and $\varepsilon=10^{-8}$, and $p_0,p_1,p_2,p_3$ are just quantities from the fits that vary for different $p$. For $p=3,5,7,9$, we have, for $(p_0,p_1,p_2,p_3)$, the following:
\begin{equation}
    \begin{split}
        p&=3:\,\,(8,0.339848,0.37,0.63), \\
        p&=5:\,\,(24,0.251115,0.41,0.71), \\
        p&=7:\,\,(48,0.682602,0.42,0.74), \\
        p&=9:\,\,(80,1.32953,0.43,0.76).
    \end{split}
    \nonumber
\end{equation}
We find that the fitted approximation does very well compared to the exact form, as is shown in Fig.~\ref{fig:Kfunc_comp}, where we plot the ratio of the exact form of $K_p$ (Eq.~\eqref{eq:K_p}) to the approximation given in Eq.~\eqref{eq:K_p_approx}, both functions of $v/w$, plotted with respect to $v/w$. The fit differs from the exact form by at most 2\%.

\bsp	
\label{lastpage}
\end{document}